\documentclass[12pt,a4paper]{article}
\usepackage[T2A]{fontenc}
\usepackage[cp1251]{inputenc}
\usepackage{cite}
\usepackage{mathtext}
\usepackage{amssymb,amsthm,amsmath}
\usepackage{array}
\usepackage[dvips]{epsfig}
\usepackage[russian, english]{babel}
\usepackage[title,titletoc]{appendix}

\begin{document}
\author{\bf Yu.A. Markov$\!\,$\thanks{e-mail:markov@icc.ru}
, M.A. Markova$\!\,$\thanks{e-mail:markova@icc.ru}}
\title{\Large \bf Mapping between the classical and pseudoclassical\\
models of a relativistic spinning particle in external\\
bosonic and fermionic fields. I}
%
%%%%%%%%%%%%%%%%%%%%%%%%%%%%%%%%%%%%%%%%%
%
\date{\normalsize \it Institute for System Dynamics and Control Theory Siberian Branch\\
of Academy of Sciences of Russia, P.O. Box 1233, 664033 Irkutsk, Russia}
\thispagestyle{empty}
\maketitle{}
%%%% Making dual numbered equations %%%%%%%%%%%%
\def\theequation{\arabic{section}.\arabic{equation}}

{
\noindent
{\bf Abstract.} The problem on mapping between two Lagrangian descriptions (using a commuting $c$-number spinor $\psi_{\alpha}$ or anticommuting pseudovector $\xi_{\mu}$ and pseudoscalar $\xi_5$ variables) of the spin degrees of freedom of a color spinning massive particle interacting with background non-Abelian gauge field, is considered. A general analysis of the mapping between a pair of {\it Majorana} spinors $(\psi_{\alpha},\hspace{0.02cm} \theta_{\alpha})$ ($\theta_{\alpha}$ is some auxiliary anticommuting spinor) and a {\it real anticommuting} tensor aggregate $(S, V_{\mu}, \!\,^{\ast}T_{\mu \nu}, A_{\mu}, P)$, is presented. A complete system of bilinear relations between the tensor quantities, is obtained. The analysis we have given is used for the above problem of the equivalence of two different ways of describing the spin degrees of freedom of the relativistic particle. The mapping of the kinetic term $(i\hbar/2)(\bar{\theta}\theta) (\dot{\bar{\psi}} \psi - \bar{\psi} \dot{\psi})$, the term  $(1/e)(\bar{\theta}\theta)\hspace{0.02cm}\dot{x}_{\mu}(\bar{\psi}\hspace{0.02cm}\gamma^{\mu}\psi)$ that provides a couple of the spinning variable $\psi$ and the particle velocity $\dot{x}_{\mu}$,  and the interaction term $\hbar\hspace{0.025cm}(\bar{\theta}\theta)\hspace{0.02cm}Q^{a\!}F_{\mu \nu}^a (\bar{\psi}\hspace{0.02cm}\sigma^{\mu \nu} \psi)$  with an external non-Abelian gauge field, are considered in detail. In the former case a corresponding system of bilinear identities including both the tensor variables and their derivatives $(\dot{S}, \dot{V}_{\mu},\!\,^{\ast}\dot{T}_{\mu \nu}, \dot{A}_{\mu}, \dot{P})$, is defined. A detailed analysis of the local bosonic symmetry of the Lagrangian with the commuting spinor $\psi_{\alpha}$, is carried out. A connection of this symmetry with the local SUSY transformation of the Lagrangian containing anticommuting pseudovector and pseudoscalar variables, is considered. The approach of obtaining a supersymmetric Lagrangian in terms of the even $\psi_{\alpha}$ and odd $\theta_{\alpha}$ spinors,  is offered.
}
{}

\newpage

%%%%%%%%%%%%%%%%%%%%%%%%%%%%%%%%%%%%%%%%%%%%%%%%%

\section{Introduction}
\setcounter{equation}{0}
\label{section_1}

In our paper \cite{markov_J_Phys_G_2010} the model Lagrangian describing the interaction of a relativistic spinning color-charged classical particle with background non-Abelian gauge and fermion fields was suggested. The spin degrees of freedom have been presented in \cite{markov_J_Phys_G_2010} by a $c$\,-number Dirac spinor $\psi_{\alpha}$, $\alpha = 1,\ldots,4$. By virtue of the fact that the background fermion field $\Psi_{\alpha}^i(x)$ (which within the classical description is considered as a Grassmann-odd function) has, by definition, the spinor index $\alpha$, the description of the spin degrees of freedom of the particle in terms of the spinor $\psi_{\alpha}$ is very natural and simplest in technical respect. There is some vagueness with respect to Grassmann parity of this spinor. In our papers \cite{markov_NPA_2007, markov_IJMPA_2010} in application to an analysis of dynamics of a spinning color particle moving in a hot quark-gluon plasma, the spinor $\psi_{\alpha}$ was thought as the Grassmann-even parity one (although it is not improbable that simultaneous using of spinors of the different Grassmann parity may be required for a complete classical description of the spin dynamics in external fields of different statistics). Furthermore, for simplicity throughout our previous works \cite{markov_NPA_2007, markov_IJMPA_2010, markov_TMF_1995, markov_TTSP_1999, markov_Ann_Phys_2005}, we have neglected a change of the spin state of the particle, i.e. we believed $\psi_{\alpha}$ to be a spinor independent of the evolution parameter $\tau$. As a result we have completely neglected an influence of the spin of particle on the general dynamics of the interaction of the particle with background fields. However, for a more detailed study of the motion of a particle in external fields of different statistics and comparing the suggested model with the other approaches known in the literature, it is necessary to account for a change in time of the spin variable $\psi_{\alpha}$. At present there exist a few approaches to the description of the spin degrees of freedom of a particle within the (semi)classical approximation. Below only two approaches closely related to the subject of our subsequent investigation are outlined.\\
\indent
Notice that the description of the spin degrees of freedom by means of a classical commuting spinor is not new. Such a way of the description arises naturally in determining the connection of relativistic quantum mechanics of an electron with relativistic classical mechanics \cite{pauli_1932}. In particu\-lar, it was shown \cite{fock_1937, akhiezer_1969, bolte_1999} that within the WKB-method extended to the relativistic case, the relativistic wave Dirac equation results in a system of equations incorporating not only the classical canonical equations of motion, but also yet another equation for the spin degrees of freedom. This equation is connected directly with the Schr\"odinger equation
\begin{equation}
i\hbar\,\frac{d\psi(\tau)}{d\tau} =
-\frac{q\hbar}{4\hspace{0.02cm}m}\,\sigma^{\mu \nu} F_{\mu \nu}(x)\psi(\tau)
\label{eq:1q}
\end{equation}
for the commuting spinor function $\psi_{\alpha}$ (we put throughout $c=1$ for the speed of light). Here, $\sigma^{\mu\nu} = [\gamma^{\mu},\gamma^{\nu}]/2i$ and $q$ is an electric charge. This equation describes the motion of the spin of the electron in a given electromagnetic field $F_{\mu \nu}(x)$. The field in (\ref{eq:1q}) is defined along the path of particle $x_{\mu} = x_{\mu} (\tau,x_0,\tau_0)$ in four-dimensional Minkowski space
(‘mostly minus’ metric), as a function of the proper time $\tau$.\\
\indent
Further, Bohm {\it et al.} \cite{bohm_1955} have introduced two-component spinor in {\it classical} (non-relativistic) hydrodynamics in view of obtaining a causal model for the Pauli equation. Their method consists in associating the spinor with the rotation of an element of the fluid. Unfortunately, this method breaks down in point mechanics and it is difficult to extend it to the relativistic case. Another line of thought is due to Proca \cite{proca_1954} who attaches a bispinor to a point particle. He then proceeds to find the equation of motion for the bispinor  which would lead to the correct equation for the velocity 4-vector and the spin antisymmetrical tensor. He has considered the case of a free point particle and the particle in an external gauge field. Based on the most general heuristic considerations he has suggested the Lagrangian, which in our notations is
\begin{equation}
L = \frac{1}{2}\,i \Lambda \biggl( \frac{d \bar{\psi}}{d \tau}\,\psi - \bar{\psi}\,\frac{d \psi} {d\tau}\biggr) +
p_{\mu} (\dot{x}^{\mu} - \bar{\psi} \gamma^{\mu} \psi ) + q A_{\mu}(x)( \bar{\psi} \gamma^{\mu} \psi ) -
\Lambda\frac{q}{4\hspace{0.02cm}m} F_{\mu \nu} (x)( \bar{\psi} \sigma^{\mu \nu} \psi ).
\label{eq:1w}
\end{equation}
Here, $A_{\mu}(x)$ is an electromagnetic four-potential, $\Lambda$ is a constant with the dimension of action (in this case the spinor $\psi_{\alpha}(\tau)$ is a dimensionless function) and the momentum $p_{\mu}$ is considered as a Lagrange multiplier for the constraint
\vspace{-0.1cm}
\begin{equation}
\dot{x}_{\mu} = \bar{\psi} \gamma_{\mu}\psi,
\label{eq:1e}
\end{equation}
where the dot denotes differentiation with respect to $\tau$. Within the framework of the classical model the whole phase space consists of the usual pair of conjugate variables $(x_{\mu}, p_{\mu})$ and of another pair of conjugate classical spinor variables $(\psi, -i \bar{\psi})$ represen\-ting the internal degree of freedom. The configuration space is thus ${\rm M}_4\otimes \mathbb{C}_4,\,\psi\in \mathbb{C}_4$ and the Lagrangian (\ref{eq:1w}) describes a symplectic system.\\
\indent
Independently of Bohm and Proca, G\"ursey \cite{gursey_1955} has developed the spinor formulation of relativistic kinematics readily applicable to a free point particle by introducing in classical theory a bispinor with a precise geometrical meaning showing its relation to the wave function of a Dirac particle. In the paper by Takabayasi \cite{takabayasi_1959} the bispinor was used for the description of general kinematical and dynamical aspects of relativistic particles possessing internal angular velocity together with internal angular momentum. In the above-mentioned papers the foundation for the theory of the proper bispinor (or simply spinor) associated with the relativistic motion of a point particle has been laid. Such a way of the description of the spin degrees of freedom of elementary particle has been used extensively by Barut with co-workers \cite{barut_1984, barut_pavsic_1987} (see also \cite{kar_2011}). They employed for this purpose the Lagrangian (\ref{eq:1w}) without the last term. The total Lagrangian  (\ref{eq:1w}) together with the last term was reproduced by Barut and Pav{\v s}i{\v c} \cite{barut_pavsic_1988} within the five dimensional Barut-Zanghi model \cite{barut_1984} treated \`a la Kaluza-Klein. Finally, we note that the model Lagrangian (\ref{eq:1w}) (in the free case) is similar to the model discussed by Plyushchay in \cite{plyushchay_1_1990}.\\
\indent
Further, in the papers \cite{berkovits_2002, berkovits_2001} by Berkovits the BRST invariant actions for a ten-dimensional superparticle moving in super-Maxwell and super-Yang-Mills backgrounds, respectively, have been written out. The equation (\ref{eq:1q}) for the commuting spinor $\psi_{\alpha}$ in \cite{berkovits_2002, berkovits_2001} is complemented by the appropriate equation for  an odd spinor variable $\theta_{\alpha}$ (Grassmann coordinate). A more detailed  discussion of Berkovits's approach will be given in the concluding section of this paper. Finally, we note that two commuting complex two-component spinor wavefunctions have been used in \cite{biedenharn_1988} for describing the massive and massless particles of half-integer spin.\\
\indent
In approach suggested in the present work we give up the constraint (\ref{eq:1e}). Next, we define the interaction term with an external non-Abelian gauge field in such a way that the term was in agreement with the equation of motion (\ref{eq:1q}). Under these circumstances we suggest the following model Lagrangian that takes into account a change both in the color and in the spinning degrees of freedom of a classical particle propaga\-ting in the background non-Abelian gauge field
\begin{equation}
L = L_0  + L_m + L_{\theta},
\label{eq:1t}
\end{equation}
where
\begin{align}
\hspace{0.6cm}
&L_0 = -\frac{1}{2\hspace{0.01cm}e}\,\bigl(\dot{x}_{\mu} - \lambda\hspace{0.02cm}(\bar{\psi}\gamma_{\mu}\psi)\bigr)^2 +\,
\hbar\hspace{0.01cm}\lambda\hspace{0.05cm}\frac{i}{2}\,\bigl(\dot{\bar{\psi}}\psi - \bar{\psi}\dot{\psi}\bigr) = \notag \\
&\hspace{0.46cm} = - \frac{1}{2\hspace{0.01cm}e}\,\dot{x}_{\mu}\dot{x}^{\mu}
+ \frac{1}{e}\,\lambda\hspace{0.02cm}\dot{x}_{\mu}(\bar{\psi}\gamma^{\mu}\psi)
-\frac{1}{2e}\,\lambda^2(\bar{\psi}\gamma_{\mu}\psi) (\bar{\psi}\gamma^{\mu}\psi) +\,
\hbar\hspace{0.01cm}\lambda\hspace{0.05cm}\frac{i}{2}\,\bigl(\dot{\bar{\psi}}\psi - \bar{\psi}\dot{\psi}\bigr), \label{eq:1y} \\
&L_m = -\frac{e}{2}\,m^2, \label{eq:1u} \\
&L_{\theta} = i\hspace{0.005cm}\hbar\hspace{0.035cm}\bigl(\theta^{\dagger\hspace{0.01cm}{i}}D^{ij}(A)
\theta^{j}\bigr) - \hbar\hspace{0.025cm}\lambda\hspace{0.01cm}\,\frac{e\hspace{0.02cm}g}{4}\,Q^{a\!}
F^{a}_{\mu\nu}(\bar{\psi}\sigma^{\mu\nu}\psi). \label{eq:1i}
\end{align}
\indent
The Lagrangian that is closely similar to the one (\ref{eq:1t}) without taking into account the interaction with an external gauge field, was discussed by Hasiewic {\it et al.} in \cite{hasiewicz_1992}. The authors have shown that in the case when $\psi$ is a commuting Majorana spinor, the quantization of this model gives a unified quantum-mechanical description for massive and massless particles of arbitrary spin and helicity. The classical Lagrangian (\ref{eq:1t}) (in the free case) was obtained from the one with so-called doubly supersymmetry \cite{kowalski-glikman_1988} after putting all Grassmann variables equal to zero and adding a kinetic term for the commuting spinor. We note that the last but one term in (\ref{eq:1y}) vanishes for the Majorana spinor $\psi$ due to the Fierz identity.\\
\indent
In (\ref{eq:1t}), in contrast to (\ref{eq:1w}), we have set $\Lambda \equiv \hbar\hspace{0.02cm}\lambda$, where $\lambda$ is some dimensionless constant, whose explicit form will be defined in the next section; $D^{ij}(A)\! =\! \delta^{ij}\partial/\partial\tau\!+\!\,i(g/\hbar)\hspace{0.02cm}\dot{x}^{\mu\!}A^{a}_{\mu}(t^{a})^{ij}$
is the covariant derivative along the direction of motion; $e$ is the (one-dimensional) vierbein field;
the self-conjugate pair $(\theta^{\dagger{i}},\hspace{0.01cm}\theta^i)$ is a set of Grassmann variables belonging to the fundamental representation of the $SU(N_c)$ color group\footnote{Here, one can draw some interesting analogy to (super)string theory for the interacting terms in (\ref{eq:1i}). In our case the term $\dot{x}^{\mu} A_{\mu}^a (\theta^{\dagger} t^a \theta)$ is similar that $j^a \bar{\partial}x^{\mu}A_{\mu}^a(x)$ defining the interaction with the so-called Neveu-Schwarz $(N\!S\hspace{0.01cm}N\!S)$ gauge fields \cite{callan_1985}. Another term of the form $F_{\mu\nu}(\bar{\psi}\sigma^{\mu\nu}\psi)$ represents analog of the term $\bar{S}\Gamma^{[\hspace{0.03cm}\mu_1 \ldots} \Gamma^{\mu_n]\!} S F_{\mu_1 \ldots \mu_n}$ for $n=2$,
where $\bar{S}_{\alpha}$ and $S_{\alpha}$ are the spin fields. The latter term in string theory defines the interaction with the Ramond-Ramond $(RR)$ gauge fields \cite{polchinski_1995}.
$N\!S\hspace{0.01cm}N\!S$ and $RR$ gauge fields are quite different in string theory in contrast to the theory of point particles.}, i.e. $i,j,\ldots =\! 1,\ldots, N_c$, (while $a,\,b,\ldots \mbox{run from}\, 1 \,\mbox{to}\, N^2_c-1$); the commuting color charge $Q^a$ is defined by
\[
\quad
Q^a\equiv\theta^{\dagger\hspace{0.01cm}{i}}(t^{a})^{ij}\theta^{j}.
\]
By virtue of the fact that we have introduced the Planck constant $\hbar$ as a factor in the first term in (\ref{eq:1i}), the color charges $\theta^{\dagger\hspace{0.01cm}{i}}(\tau)$ and $\theta^{i}(\tau)$ are dimensionless variables like the spinor function $\psi_{\alpha}(\tau)$. Besides, it is worthy of special emphasis that we kept $\hbar$ in denominator in the second term of the covariant derivative $D^{ij}(A)$ (see the definition above), as is generally accepted in the field theory. In this case the group generators $t^a$ are dimensionless quantities and the non-Abelian gauge field $A_{\mu}^a(x)$ has the canonical dimension. The disadvantage of such an approach is that $\hbar$ will enter in an explicit form into the classical equations of motion for the color charges $\theta^i$ and $Q^a$, Eqs.\,(\ref{ap:A9}), (\ref{ap:A11}), and also into the covariant derivative $D_{\mu}^{a b}(A)$ in the Lorentz equation (\ref{ap:A10}) and the Yang-Mills equation. Here we follow the papers by Arodz \cite{arodz_1_1982, arodz_2_1982}. Recall that in the original paper by Wong \cite{wong_1970} the group generators $t^a$ are dimensional quantities, i.e. $t^a \equiv \frac{1}{2} \hbar\hspace{0.045cm} \lambda^a$, where $\lambda^a$ are the Gell-Mann matrices and thus
\[
[\hspace{0.02cm}t^a, t^b\hspace{0.02cm}] = i\hbar f^{abc} t^c.
\]
In Wong's approach the classical color charge $Q^a$ was identified with the expectation value of the operator $\hat{t}^a = \frac{1}{2} \hbar\hspace{0.02cm} \lambda^a$, i.e. $Q^a \equiv \langle\hspace{0.02cm}\hat{t}^a\rangle$, by analogy with the spin vector
$\mathbf{S} \equiv \, \langle\frac{1}{2} \hbar\hspace{0.02cm} \hat{\boldsymbol{\sigma}}\rangle$. Such point of view was also accepted in a number of the subsequent papers concerning a given subject (see, for example, \cite{heinz_1984, guhr_2007}). In this case the Planck constant $\hbar$ disappears in the equations of motion (A.9), (A.11) etc. However the dimension of gauge field in this case will not already be the canonical one as it is accepted in the field theory\footnote{\,In the paper \cite{guhr_2007} a qualitative argument on this matter has been given. If we take $t^{a} \sim \hbar$ and at the same time believe that a gauge field is of order $1/\hbar$,
i.e.
\[
A_{\mu}^a(x) \sim 1/\hbar,
\]
then this leads to the fact that the interaction of a color particle with the non-Abelian gauge field has the same $\hbar$-dependence as in quantum electrodynamics. In QCD such a strong field is called an {\it external} color field. Just that case we mean in the present paper. However, in the situation which was accepted in \cite{wong_1970, heinz_1984, guhr_2007} the original field $A_{\mu}^a(x)$ is of order 1. Such the color fields are {\it microscopic} or {\it dynamical} fields, that falls into purely quantum branch.}.\\
\indent
The alternative approach most generally employed for the description of spin for a massive point particle is connected with introduction into consideration of the pseudovector and pseudoscalar dynami\-cal
variables $\xi_{\mu}$ and $\xi_5$ that are elements of the Grassmann algebra \cite{berezin_1975, barducci_1976, brink_1976, balachandran_1977}. For these variables an appropriate Lagrangian of the first order time derivative, was defined. In view of its great importance for a further discussion and for convenience of future references we give in Appendix A a complete form of this Lagrangian. It is these Grassmann-valued variables that appear in the representation of the one-loop effective action in quantum chromodynamics in terms of the path integral over world lines of a hard particle moving in external non-Abelian gauge field \cite{borisov_1982, strassler_1992, d_hoker_1996}. \\
%
%We notice also that the $\xi_{\mu}$ variable in two-dimensional case is used in the description of a spinning string within the Ramond-Neveu-Schwarz formalism, and in particular in the construction of the covariant string vertex operators  describing emission (absorption) of gauge vector particles \cite{andreev_1988} and fermions \cite{knizhnik_1985, friedan_1986} by the string.\\
\indent
The description of the spin degrees of freedom in terms of the odd pseudovector and pseudosca\-lar quantities $(\xi_{\mu}, \xi_5)$ is to some extent more fundamental in comparison with the description in terms of the commuting spinor $\psi_{\alpha}$. For this reason the interesting question arises as to whether it is possible to define a mapping between these variables, and, finally, the possibility of constructing the mapping between the Lagrangians (\ref{eq:1t}) and (\ref{ap:A1}). The mapping of this type was first considered by Barut and Pav{\v s}i{\v c}  \cite{barut_pavsic_1989} (see also Scholtz {\it et al.} \cite{scholtz_1995}).\\
\indent
It is pertinent at this point to make one remark which is completely analogous to that made in Introduction of the paper \cite{markov_J_Phys_G_2010}. This remark was concerned with introducing into consideration the Grassmann color charges $\theta^{\dagger{i}}$ and $\theta^{i}$. If we closely look at the equations of motion (\ref{ap:A9})\,--\,(\ref{ap:A11}), and at the expression for the color current (\ref{ap:A12}), then we may observe that the odd pseudovector $\xi_{\mu}$ enters these equations only in the following {\it even} combination
{\setlength\abovedisplayskip{7pt plus 0pt minus 2pt}
\setlength\belowdisplayskip{7pt plus 0pt minus 2pt}
\begin{equation}
S^{\mu\nu}\!\equiv -i\hspace{0.04cm}\xi^{\mu}\xi^{\nu},
\label{eq:1o}
\end{equation}
as well as the Grassmann color charges enter these equations in the combination $\theta^{\dagger}t^a\theta\,(\equiv Q^a)$. By virtue of (\ref{ap:A8}) the function $S^{\mu \nu}$
obeys the equation of motion
\begin{equation}
\frac{dS^{\mu\nu}}{d\tau}=\frac{g}{m}\,Q^a(F^{a\hspace{0.02cm}\mu}_{\;\;\;\;\,\lambda}S^{\lambda\nu} - F^{a\hspace{0.02cm}\nu}_{\;\;\;\;\,\lambda}S^{\lambda\mu}).
\label{eq:1p}
\end{equation}
One notices that a similar tensor of spin can be defined also in terms of the $\psi_{\alpha}$ spinor
\begin{equation}
S^{\mu\nu} = \frac{1}{2}\hspace{0.04cm}\hbar\hspace{0.03cm}\lambda(\bar{\psi}\sigma^{\mu\nu}\psi).
\label{eq:1a}
\end{equation}
By virtue of (\ref{eq:1q}) this tensor of spin obeys the same equation (\ref{eq:1p}).\\
\indent
Thus in the actual dynamics of a classical color spinning particle one may quite manage with the usual commuting function $S^{\mu\nu}$. The odd pseudovector variable $\xi^{\mu}$ gives merely the possibility of a classical Lagrangian formulation without any dynamical effects. One can expect that the situation can qualitatively change only if the system is subjected to background (non-)Abelian fermion field which as it were `splits' the combination $S_{\mu\nu}\!=\!-i\hspace{0.02cm}\xi_{\mu\,}\xi_{\nu}$ into two independent Grassmann-odd parts (see our second part \cite{markov_part_II}). Here, the necessity of introducing the Grassmann pseudovector $\xi_{\mu}$ as a dynamical variable enjoying full rights should be manifested in full.\\
\indent
Further, by virtue of the fact that we have the even spinor $\psi_{\alpha}$ on the one hand and the odd pseudovector $\xi_{\mu}$ (and pseudoscalar $\xi_5$) on the other hand, for the construction of the desired mapping we must introduce some auxiliary odd spinor $\theta_{\alpha}$. The idea of the construction of such a mapping is not new. In due time this problem has been studied extensively in view of analysis of a classical correspondence of theories of relativistic massless spin-1/2 particles \cite{berezin_1975, barducci_1976, brink_1976} and superparticles \cite{casalbuoni_1976, ferber_1978, volkov_1980, brink_1_1981, brink_2_1981} and in a more general context between spinning strings and superstrings. In the paper by Sorokin {\it et al.} \cite{sorokin_1989} within the superfield formalism it was noted that such a classical correspondence can be defined by the following relation:
\begin{equation}
\xi_{\mu} \sim \bar{\theta}\gamma_{\mu}\psi + (\mbox{conj.\,part}).
\label{eq:1d}
\end{equation}
In \cite{sorokin_1989} the commuting spinor $\psi_{\alpha}$ played the role of a twistor-like variable which is not dynamical one. In our paper we use the relation (\ref{eq:1d}). The only difference is that by virtue of initial statement of the problem, the anticommuting spinor $\theta_{\alpha}$ will play a role of the auxiliary variable rather than $\psi_{\alpha}$.\\
\indent
The paper is organized as follows. In Section \ref{section_2} a general analysis of the mapping between a pair of the Dirac spinors $(\psi_{\alpha}, \, \theta_{\alpha})$ and the {\it real anticommuting} tensor system $(S, V_{\mu},\!\,^{\ast}T_{\mu \nu}, A_{\mu}, P)$ is proposed. The required initial general expressions of the mapping are written out. An important special case when the spinors $\psi_{\alpha}$ and $\theta_{\alpha}$ are the Majorana ones, is considered. The algebraic relations between the tensor quantities are defined with the help of the Fierz identities.
Section \ref{section_3} is devoted to the discussion of mapping the kinetic term for the commuting spinor $\psi$ in (\ref{eq:1y}). Here, the required general expression connecting the kinetic term with the derivative of the tensor quantities $(\dot{S}, \dot{V}_{\mu}, \!\,^{\ast}\dot{T}_{\mu \nu}, \dot{A}_{\mu}, \dot{P})$, is defined. A limiting case of the Majorana spinors is considered. The procedure of deriving the algebraic relations including aside from the tensor quantities itself, also their derivatives, is described  in full. In Section \ref{section_4} a detailed analysis of the local bosonic symmetry of the free Lagrangian (\ref{eq:1t}) is carried out and a connection of this symmetry with the reduced local supersymmetry transformations presented in Appendix A, is considered. In Section \ref{section_9} a qualitative consideration of supersymmetric generalization of our initial Lagrangian (\ref{eq:1t}) is performed. In the concluding Section \ref{section_12} we briefly discuss the question of further generalization of the ideas of this work, namely, the generalization of the original classical Lagrangian (\ref{eq:1t}) to the supersymmetric case.\\
\indent
In Appendix A a complete form of the Lagrangian for a spin-$\frac{1}{2}$ color particle is given and the local SUSY $n=1$ transformations, constraints and equations of motion are written out. In Appendix B all of the necessary formulas of spinor algebra are listed. In Appendix C a complete list of all 15 sets of the bilinear relations connecting the real currents $(S, V_{\mu}, \!\,^{\ast}T_{\mu \nu}, A_{\mu}, P)$  among themselves is set out. The above-mentioned list is introduced in such a way that it simultaneously covers both the commutative and anticommutative cases of the current variables.\\
\indent
In Appendix D it is given the proof of independence of mapping the kinetic term for spinor variable (\ref{eq:1y}) from the fact whether the auxiliary term $\theta_{\alpha}$ is constant or variable quantity, provided the commuting spinor $\psi_{\alpha}$ and anticommuting auxiliary spinor $\theta_{\alpha}$ are the Majorana ones.

%%%%%%%%%%%%%%%%%%%%%%% section 2 %%%%%%%%%%%%%%%%%%%%

\section{\bf General analysis of a connection between a pair of spinors $(\psi_{\alpha},\theta_{\alpha})$ and anticommuting tensor system}
\setcounter{equation}{0}
\label{section_2}

The problem of defining a mapping between the commuting spinor $\psi_{\alpha}$ and anticommuting pseudovector and pseudoscalar variables $(\xi_{\mu},\,\xi_5)$ is in fact a part of a more general analysis of the relation between spinors (Dirac, Majorana or Weyl ones) and Lorentz-invariant real or complex tensor systems. In the case of a commuting c-number Dirac spinor and 16 real commuting bilinear tensor quantities that are formed by the given spinor, such a problem has been studied by Takahashi and Okuda \cite{takahashi_1982, takahashi_1983}, Kaempffer \cite{kaempffer_1981}, Crawford \cite{crawford_1985}, Lounesto \cite{lounesto_1993} and from a somewhat different viewpoint by Zhelnorovich \cite{zhelnorovich_1970, zhelnorovich_1972, zhelnorovich_book}. The latter author has also considered  the special cases of Majorana and Weyl spinors, and the most important for us problem of the relation between a pair of {\it two commuting} spinors $(\psi_{\alpha},\varphi_{\alpha})$ and appropriate tensor set\footnote{\,It seems likely that the decomposition of the direct product of two commuting spinors in terms of tensors was first discussed by Case \cite{case_1955} (see also \cite{corson_1955, hamilton_1984}).}. In the subsequent discussion we will follow essentially Zhelnorovich \cite{zhelnorovich_1972, zhelnorovich_book}.
At the end of this section we will mention another alternative approach based on the K\"ahler formalism \cite{kahler_1962}, which represents fermions in terms of antisymmetric tensor fields.
\\
\indent
In the problem under consideration we also have at hand two, in the general case Dirac spinors $\psi_{\alpha}$ and $\theta_{\alpha}$ (although, as was mentioned in Introduction, in our case the latter plays an auxiliary role). However, the second spinor is now classical {\it anticommuting} one as distinct from the works \cite{zhelnorovich_1972, zhelnorovich_book}.\\
\indent
The heart of our subsequent considerations is the expansion of the spinor structure $\hbar^{1/2\,}\bar{\theta}_{\beta} \psi_{\alpha}$ in the basis of the Dirac $\gamma$-matrices:
\begin{equation}
\hbar^{1/2\,}\bar{\theta}_{\beta}\psi_{\alpha} = \frac{1}{4}\,
\Bigl\{-i\hspace{0.02cm}S\hspace{0.01cm}\delta_{\alpha\beta} + V_{\mu}(\gamma^{\mu})_{\alpha\beta} - \frac{i}{2}\,^{\ast}T_{\mu\nu}(\sigma^{\mu\nu}\gamma_{5})_{\alpha\beta} +
i\hspace{0.01cm}A_{\mu}(\gamma^{\mu}\gamma_{5})_{\alpha\beta} + P(\gamma_{5})_{\alpha\beta}\Bigr\}.
\label{eq:2q}
\end{equation}
The expansion for the Hermitian conjugate expression is
\begin{equation}
\hbar^{1/2\,}\bar{\psi}_{\beta}\theta_{\alpha} = \frac{1}{4}\,
\Bigl\{i\hspace{0.02cm}S^{\hspace{0.02cm}\ast}\hspace{0.01cm}\delta_{\alpha\beta} + V_{\mu}^{\ast}(\gamma^{\mu})_{\alpha\beta} -
\frac{i}{2}\,(\,^{\ast}T_{\mu\nu})^{\ast}(\sigma^{\mu\nu}\gamma_{5})_{\alpha\beta} -
i\hspace{0.01cm}A_{\mu}^{\ast}(\gamma^{\mu}\gamma_{5})_{\alpha\beta} - P^{\hspace{0.02cm}\ast}(\gamma_{5})_{\alpha\beta}\Bigr\}.
\label{eq:2w}
\end{equation}
Here, the  {\it complex anticommuting} tensor variables on the right-hand side are defined as follows:
\begin{equation}
\begin{split}
S \equiv i\hspace{0.02cm}\hbar^{1/2}&(\bar{\theta}\hspace{0.02cm}\psi),\quad\;  V_{\mu} \equiv \hbar^{1/2}(\bar{\theta}\gamma_{\mu}\hspace{0.02cm}\psi),\quad\;
\,^{\ast}T_{\mu\nu} \equiv i\hspace{0.02cm}\hbar^{1/2}(\bar{\theta}\sigma_{\mu\nu}\gamma_{5\hspace{0.02cm}}\psi),\quad\;\\
&A_{\mu} \equiv i\hspace{0.015cm}\hbar^{1/2}(\bar{\theta}\gamma_{\mu}\gamma_{5\hspace{0.02cm}}\psi),\quad\; P \equiv \hbar^{1/2\,}(\bar{\theta}\gamma_{5\hspace{0.02cm}}\psi).\\
\end{split}
\label{eq:2e}
\end{equation}
The multiplies on the right-hand side of expressions in (\ref{eq:2e}) have been chosen such that for {\it Majorana} spinors $\psi_{\alpha}$ and $\theta_{\alpha}$ the tensor quantities (\ref{eq:2e})
are real, i.e.
\begin{equation}
S = S^{\hspace{0.02cm}\ast}, \quad V_{\mu} = V_{\mu}^{\ast}, \quad \,^{\ast}T_{\mu\nu} = (\,^{\ast}T_{\mu\nu})^{\ast},\quad \ldots\;.
\label{eq:2r}
\end{equation}
The symbol $\ast$ denotes complex conjugation. On the left-hand side of the expressions (\ref{eq:2q}) and (\ref{eq:2w}) we have introduced the dimensional factor $\hbar^{1/2}$, since the spinors $\psi_{\alpha}$ and $\theta_{\alpha}$ are considered as dimensionless variables (see Introduction), while the dimension of the functions $S$, $V_{\mu}$, $\!\,^{\ast}T_{\mu\nu}$, etc. is $[S]\sim [V_{\mu}] \sim [\,^{\ast}T_{\mu\nu}]\sim\ldots\sim [\hbar\,]^{1/2}$.\\
\indent
Write out next an important formula for the product of two expressions (\ref{eq:2q}) and (\ref{eq:2w})
\begin{equation}
\hbar\hspace{0.025cm}(\bar{\theta}_{\beta}\psi_{\alpha})(\bar{\psi}_{\gamma}\theta_{\delta}) =
\label{eq:2t}
\end{equation}
\[
\begin{split}
\frac{1}{16}\,
&\Bigl\{-i\hspace{0.02cm}S\hspace{0.02cm}\delta_{\alpha\beta} + V_{\mu}\hspace{0.01cm}(\gamma^{\mu})_{\alpha\beta} - \frac{i}{2}\,^{\ast}T_{\mu\nu}(\sigma^{\mu\nu}\gamma_{5})_{\alpha\beta} +
i\hspace{0.01cm}A_{\mu}(\gamma^{\mu}\gamma_{5})_{\alpha\beta} + P(\gamma_{5})_{\alpha\beta}\Bigr\}\times\\
\times\, &\Bigl\{i\hspace{0.025cm}S^{\hspace{0.02cm}\ast}\hspace{0.01cm}\delta_{\delta\gamma} + V_{\nu}^{\ast}(\gamma^{\nu})_{\delta\gamma} -
\frac{i}{2}\,(\,^{\ast}T_{\lambda\sigma})^{\ast}(\sigma^{\lambda\sigma}\gamma_{5})_{\delta\gamma} -
i\hspace{0.01cm}A_{\nu}^{\ast}(\gamma^{\nu}\gamma_{5})_{\delta\gamma} - P^{\ast}(\gamma_{5})_{\delta\gamma}\Bigr\}.
\end{split}
\]
\indent
Let us consider for example the simplest crossed contraction of the expression (\ref{eq:2t}) with  $\delta_{\beta \delta} \delta_{\gamma \alpha}$. As a result we obtain
\begin{equation}
\hbar\hspace{0.02cm}(\bar{\theta}\theta)(\bar{\psi}\psi) = \frac{1}{4}\Bigl\{SS^{\hspace{0.02cm}\ast} + V_{\mu}(V^{\mu})^{\ast} - \frac{1}{2}\,^{\ast}T_{\mu\nu}(\,^{\ast}T^{\mu\nu})^{\ast} -
A_{\mu}(A^{\mu})^{\ast} - PP^{\hspace{0.02cm}\ast}\Bigr\}.
\label{eq:2y}
\end{equation}
Let the commuting spinor $\psi$ and the anticommuting spinor $\theta$ be Majorana (M) ones. As is known in this case the following relations hold
\begin{equation}
\begin{split}
&(\bar{\psi}_{\rm M}\hspace{0.02cm}\psi_{\rm M}) = 0,\quad\;  (\bar{\psi}_{\rm M}\gamma_{5\hspace{0.02cm}}\psi_{\rm M}) = 0, \quad\;
(\bar{\psi}_{\rm M}\gamma_{\mu}\gamma_{5\hspace{0.02cm}}\psi_{\rm M}) = 0,\\
&(\bar{\theta}_{\rm M}\gamma_{\mu}\hspace{0.02cm}\theta_{\rm M}) = 0, \;\;\, (\bar{\theta}_{\rm M}\sigma_{\mu\nu}\theta_{\rm M}) = 0.\\
\end{split}
\label{eq:2u}
\end{equation}
By virtue of (\ref{eq:2u}), the left-hand side of (\ref{eq:2y}) vanishes. The right-hand side of this expression is equal to zero by the condition (\ref{eq:2r}) and by nilpotency of the tensor quantities.\\
\indent
A more nontrivial expression can be obtained from (\ref{eq:2t}) by its contracting with
$\delta_{\beta \delta} (\sigma^{\mu \nu})_{\gamma \alpha}$. The calculation of the traces of the product of $\gamma$-matrices by employing the formulae in Appendix B leads to the expression
\begin{equation}
\begin{split}
\hbar&\hspace{0.02cm}(\bar{\theta}\theta)(\bar{\psi}\sigma^{\mu\nu}\psi) = \frac{i}{4}\,\Bigl\{-\bigl[\hspace{0.02cm}S(\hspace{0.02cm}T^{\mu\nu})^{\ast} - T^{\mu\nu} S^{\hspace{0.02cm}\ast}\bigr]
+ \bigl[\,^{\ast}T^{\mu\nu} P^{\hspace{0.02cm}\ast} - P (\,^{\ast}T^{\mu\nu})^{\ast}\bigr] \,+\\
&+\bigl[\hspace{0.03cm}V^{\mu}(V^{\nu})^{\ast} - V^{\nu}(V^{\mu})^{\ast}\bigr] - \bigl[\hspace{0.03cm}A^{\mu}(A^{\nu})^{\ast} - A^{\nu}(A^{\mu})^{\ast}\bigr] -
\epsilon^{\hspace{0.02cm}\mu\nu\lambda\sigma}\bigl[\hspace{0.03cm}V_{\lambda}(A_{\sigma})^{\ast} + A_{\lambda}(V_{\sigma})^{\ast}\bigr]\,+\\
&+\bigl[\,^{\ast}T^{\mu\lambda}(\,^{\ast}T_{\lambda\;\;}^{\;\;\nu})^{\ast} - \,^{\ast}T^{\nu\lambda}(\,^{\ast}T_{\lambda\;\;}^{\;\;\mu})^{\ast}\bigr]\Bigr\}.
\end{split}
\label{eq:2i}
\end{equation}
Hereafter, we make use of the formulae for going over from an arbitrary antisymmetric tensor of second rank to its dual tensor and conversely,
\[
\,^{\ast}T^{\mu\nu} = \frac{1}{2}\,\epsilon^{\hspace{0.02cm}\mu\nu\lambda\sigma}T_{\lambda\sigma}, \qquad T^{\mu\nu} = -\frac{1}{2}\,\epsilon^{\hspace{0.02cm}\mu\nu\lambda\sigma}\,^{\ast}T_{\lambda\sigma}.
\]
The expression (\ref{eq:2i}) is just the one entered in the Lagrangian (\ref{eq:1i}) for the specific
choice\footnote{\,In our choice of the constant $\lambda$, the definition of the tensor spin (\ref{eq:1a}) takes the form
\[
S^{\mu \nu}\! = \frac{1}{2}\,\hbar\hspace{0.04cm}(\bar{\theta} \theta)(\bar{\psi}\sigma^{\mu\nu}\psi).
\]
In this case only the expression acquires the nilpotency property: the product of its any five components equals zero, i.e. $\prod\limits_{i=1}^5 S^{\mu_i \nu_i} = 0$, as it takes place for the definition $S^{\mu\nu}\!\equiv -i\hspace{0.04cm}\xi^{\mu}\xi^{\nu}$.}
\begin{equation}
\lambda \equiv (\bar{\theta}\theta).
\label{eq:2ii}
\end{equation}
In the special case of Majorana spinors the expression (\ref{eq:2i}) turns to
\begin{equation}
\hbar\hspace{0.02cm}(\bar{\theta}_{\rm M}\theta_{\rm M})(\bar{\psi}_{\rm M}\sigma^{\mu\nu\!}\psi_{\rm M}) =\! \frac{i}{2}\,\Bigl\{-\!\bigl[\hspace{0.02cm}S\hspace{0.04cm}T^{\mu\nu}+ P\,^{\ast}T^{\mu\nu}\bigr]
+\bigl[\hspace{0.02cm}V^{\mu}V^{\nu} - A^{\mu}A^{\nu}\bigr] -\, \epsilon^{\mu\nu\lambda\sigma}V_{\lambda}A_{\sigma} + \!\,^{\ast}T^{\mu\lambda}\,^{\ast}T_{\lambda\;\;}^{\;\;\nu}\!\Bigr\}.
\label{eq:2o}
\end{equation}
\indent
It is very important as a self-check to consider the contraction of (\ref{eq:2t}) with a similar structure $(\sigma^{\mu \nu})_{\beta \delta} \delta_{\gamma \alpha}$. Here, instead of (\ref{eq:2i}), we have
\[
\begin{split}
\hbar&\hspace{0.02cm}(\bar{\psi}\psi)(\bar{\theta}\sigma^{\mu\nu}\theta) = \frac{i}{4}\,\Bigl\{-\bigl[\hspace{0.02cm}S(\hspace{0.02cm}T^{\mu\nu})^{\ast} - T^{\mu\nu} S^{\hspace{0.02cm}\ast}\bigr]
+ \bigl[\,^{\ast}T^{\mu\nu} P^{\hspace{0.02cm}\ast} - P (\,^{\ast}T^{\mu\nu})^{\ast}\bigr]\, -\\
&-\bigl[\hspace{0.02cm}V^{\mu}(V^{\nu})^{\ast} - V^{\nu}(V^{\mu})^{\ast}\bigr] + \bigl[A^{\mu}(A^{\nu})^{\ast} - A^{\nu}(A^{\mu})^{\ast}\bigr] -
\epsilon^{\hspace{0.02cm}\mu\nu\lambda\sigma}\bigl[V_{\lambda}(A_{\sigma})^{\ast} + A_{\lambda}(V_{\sigma})^{\ast}\bigr]\,-\\
&-\bigl[\,^{\ast}T^{\mu\lambda}(\,^{\ast}T_{\lambda\;\;}^{\;\;\nu})^{\ast} - \,^{\ast}T^{\nu\lambda}(\,^{\ast}T_{\lambda\;\;}^{\;\;\mu})^{\ast}\bigr]\Bigr\},
\end{split}
\]
and in particular for the Majorana spinors, instead of (\ref{eq:2o}), we get
\begin{equation}
\hbar\hspace{0.02cm}(\bar{\psi}_{\rm M}\psi_{\rm M})(\bar{\theta}_{\rm M}\sigma^{\mu\nu}\theta_{\rm M}) = \frac{i}{2}\,\Bigl\{-\!\bigl[\hspace{0.02cm}S\hspace{0.02cm}T^{\mu\nu}\!+ P\,^{\ast}T^{\mu\nu}\bigr]
-\bigl[\hspace{0.02cm}V^{\mu}V^{\nu} - A^{\mu}A^{\nu}\bigr] - \epsilon^{\mu\nu\lambda\sigma}V_{\lambda}A_{\sigma} - \!\,^{\ast}T^{\mu\lambda}\,^{\ast}T_{\lambda\;\;}^{\;\;\nu}\Bigr\}.
\label{eq:2p}
\end{equation}
However, here by virtue of (\ref{eq:2u}) in contrast to (\ref{eq:2o}), the left-hand side equals zero\footnote{\,The fact that the tensor expression for Majorana spinors turns to zero can be considered as indirect evidence for choosing the {\it commuting} spinor $\psi_{\alpha}$ (rather than anticommuting $\theta_{\alpha}$) as a basic dynamical variable for the classical description of the spin degrees of freedom of a particle that, generally speaking, a priory it is not at all obvious (see Introduction).} while the right-hand side represents a rather complicated algebraic expression. As distinct from (\ref{eq:2y}) it is not evident in advance that it must be also equal to zero. The proof of this fact serves a good test for correctness of a system of algebraic equations connecting the tensor quantities $S, V_{\mu}, \!\,^{\ast}T_{\mu\nu}\ldots$ among themselves (see below).\\
\indent
Not all of the quantities (\ref{eq:2e}) are independent. There exist certain algebraic relations between them. As is known, such relations are provided by the Fierz identities. According to Zhelnorovich \cite{zhelnorovich_1972, zhelnorovich_book}, the required bilinear equations can be obtained by multiplying out two expansions (\ref{eq:2q}) written for indices $\alpha, \beta$ and $\gamma, \delta$ correspondingly, followed by contracting the obtained expression with all possible bilinear products of 16 independent generators of the Clifford algebra
\[
\delta_{\beta \gamma} \delta_{\delta \alpha}, \quad \delta_{\beta \gamma} (\gamma^{\mu})_{\delta \alpha}, \quad \delta_{\beta \gamma} (\gamma_5)_{\delta \alpha}, \quad
\delta_{\beta \gamma} (\gamma^{\mu} \gamma_5)_{\delta \alpha}, \quad \delta_{\beta \gamma} (\sigma^{\mu \nu}\gamma_5)_{\delta \alpha}
\]
and so on. A full list of the algebraic equations is given in Appendix C. This system is valid for both Majorana spinors and Dirac spinors. Recall that in the latter case the tensor quantities (\ref{eq:2e})
are complex. Nevertheless the system  (\ref{ap:C1})\,--\,(\ref{ap:C15}) is unsuitable for analysis of the right-hand side of expression (\ref{eq:2i}), since here in the general Dirac case we have a product of tensor quantities (\ref{eq:2e}) and its complex conjugation. By virtue of this fact we restrict our consideration here only to the important special case of Majorana spinors (real currents (\ref{eq:2e})), when (\ref{eq:2i}) goes over into the simpler relation (\ref{eq:2o}). We shall discuss much more difficult case of Dirac spinors in Part II \cite{markov_part_II}.\\
\indent
We need equations  (\ref{ap:C3}), (\ref{ap:C6}), (\ref{ap:C8}), (\ref{ap:C10}), (\ref{ap:C12}) and (\ref{ap:C13}). For the case of anticommuting currents the equations take the form (we give these equations in the different order)
\[
\begin{split}
&S\,^{\hspace{0.02cm}\ast}T^{\mu\nu} = -\frac{1}{4}\,\epsilon^{\hspace{0.02cm}\mu\nu\lambda\sigma}V_{\lambda}V_{\sigma} - \frac{1}{4}\,\epsilon^{\hspace{0.02cm}\mu\nu\lambda\sigma}A_{\lambda}A_{\sigma} -
\frac{1}{4}\,\epsilon^{\hspace{0.02cm}\mu\nu\lambda\sigma}\,^{\ast}T_{\lambda\rho}
\,^{\ast}T^{\rho\;\;}_{\;\;\sigma},\\
&P\,^{\hspace{0.02cm}\ast}T^{\mu\nu} = \frac{1}{2}\,V^{\mu}V^{\nu} + \frac{1}{2}\,A^{\mu\!}A^{\nu} - \frac{1}{2}\,^{\ast}T^{\mu\lambda}\,^{\ast}T_{\lambda\;\;}^{\;\;\nu},\\
&V^{\mu}V^{\nu} = -\frac{1}{2}\,S\hspace{0.04cm} T^{\mu\nu} + \frac{1}{2}\,P\!\,^{\,\ast}T^{\mu\nu} - \frac{1}{2}\,\epsilon^{\hspace{0.02cm}\mu\nu\lambda\sigma}V_{\lambda}A_{\sigma},\\
&A^{\mu\!}A^{\nu} = -\frac{1}{2}\,S\hspace{0.04cm} T^{\mu\nu} + \frac{1}{2}\,P\!\,^{\,\ast}T^{\mu\nu} + \frac{1}{2}\,\epsilon^{\hspace{0.02cm}\mu\nu\lambda\sigma}V_{\lambda}A_{\sigma},\\
&V^{\mu\!}A^{\nu} = -\frac{1}{2}\,g^{\mu\nu} S\!\hspace{0.04cm}P + \frac{1}{4}\,\epsilon^{\hspace{0.02cm}\mu\nu\lambda\sigma}V_{\lambda}V_{\sigma} - \frac{1}{4}\,A^{\mu\!}A^{\nu}
+ \frac{1}{4}\,\bigl(\,^{\ast}T^{\mu\lambda}\hspace{0.02cm}T_{\lambda\;\;}^{\;\;\nu} - T^{\mu\lambda}\,^{\ast}T_{\lambda\;\;}^{\;\;\nu}\bigr).
\end{split}
\]
Here, we have omitted equation (\ref{ap:C10}) because of the awkwardness. The simple analysis of this system has shown that only three equations are independent and they can be chosen in the following form:
\begin{equation}
P\,^{\hspace{0.02cm}\ast}T^{\mu\nu} - S\hspace{0.04cm}T^{\mu\nu} = V^{\mu}V^{\nu} + A^{\mu\!}A^{\nu},
\label{eq:2a}
\end{equation}
\begin{equation}
P\,^{\hspace{0.01cm}\ast}T^{\mu\nu} + S\hspace{0.04cm}T^{\mu\nu} = -\,^{\ast}T^{\mu\lambda}\,^{\ast}T_{\lambda\;\;}^{\;\;\nu},
\hspace{0.45cm}
\label{eq:2s}
\end{equation}
\begin{equation}
-\hspace{0.02cm}\epsilon^{\hspace{0.02cm}\mu\nu\lambda\sigma}V_{\lambda}A_{\sigma} = V^{\mu}V^{\nu} - A^{\mu\!}A^{\nu}.
\hspace{0.5cm}
\label{eq:2d}
\end{equation}
By using (\ref{eq:2s}) and (\ref{eq:2d}), we obtain instead of (\ref{eq:2o})
\[
\hbar\hspace{0.02cm}(\bar{\theta}_{\rm M}\theta_{\rm M})(\bar{\psi}_{\rm M}\hspace{0.02cm}\sigma^{\mu\nu\!\hspace{0.02cm}}\psi_{\rm M}) = i\,\Bigl\{-\!\hspace{0.03cm}\bigl[\hspace{0.02cm}S\hspace{0.04cm}T^{\mu\nu}+ P\,^{\ast}T^{\mu\nu}\bigr]
+\bigl[\hspace{0.02cm}V^{\mu}V^{\nu} - A^{\mu\!}A^{\nu}\bigr] \Bigr\}.
\]
The remaining equation (\ref{eq:2a}) enables us to get rid either of $(V^{\mu} V^{\nu} - P\, ^{\ast}T^{\mu \nu})$ or of $(A^{\mu\!}A^{\nu} + S\hspace{0.04cm}T^{\mu \nu})$. In the first case we have
\begin{equation}
\hbar\hspace{0.02cm}(\bar{\theta}_{\rm M}\theta_{\rm M})(\bar{\psi}_{\rm M}\hspace{0.02cm}\sigma^{\mu\nu\!\hspace{0.02cm}}\psi_{\rm M}) = -2\hspace{0.03cm}i\bigl\{A^{\mu\!}A^{\nu} + S\hspace{0.04cm}T^{\mu\nu}\bigr\}.
\label{eq:2f}
\end{equation}
The last expression is the most important result of this section. From a comparison of the force term (\ref{eq:1i}) in different representations
\[
-\hbar\hspace{0.03cm}(\bar{\theta}\theta)\hspace{0.03cm}\frac{eg}{4}\,Q^{a\!}F^{a}_{\mu\nu}
(\bar{\psi}\hspace{0.02cm}\sigma^{\mu\nu}\psi)\sim
\frac{i\hspace{0.01cm}eg}{2}\,Q^{a\!}F^{a}_{\mu\nu\,}\xi^{\mu}\xi^{\nu} +\, \ldots\,,
\]
it follows at once that the relation
\begin{equation}
\hbar\hspace{0.02cm}(\bar{\theta}_{\rm M}\theta_{\rm M})(\bar{\psi}_{\rm M}\hspace{0.02cm}\sigma^{\mu\nu\!\hspace{0.03cm}}\psi_{\rm M}) = -2\hspace{0.02cm}i\hspace{0.03cm}\xi^{\mu}\xi^{\nu}
\label{eq:2g}
\end{equation}
must hold. A comparison with (\ref{eq:2f}) shows that we must set
\begin{equation}
A^{\mu} = \pm\, \xi^{\mu}.
\label{eq:2h}
\end{equation}
In this case for the first term on the right-hand side of (\ref{eq:2f}) we have the ideal coincidence with (\ref{eq:2g}). The second term here can be put equal to zero in the case\footnote{\label{foot_6}\,At this stage, however, we lose a possibility to construct one-to-one correspondence between two classical Lagrangians (\ref{eq:1t}) and (\ref{ap:A1}). To achieve the one-to-one correspondence we must add additional dynamic variables to the Lagrangian (\ref{ap:A1}) (or to its reduced form (\ref{ap:A14}), (\ref{ap:A15})).} when $S=0$ or $T^{\mu\nu}=0$. Further, by using equations (\ref{eq:2s}) and (\ref{eq:2d}) it is not difficult to verify that the right-hand side of (\ref{eq:2p}) vanishes, as it should be.\\
\indent
In conclusion of this section we note that among the tensor structures of the type
\[
\hbar\hspace{0.02cm}(\bar{\theta}\theta)(\bar{\psi}\Gamma^A\psi),\quad \Gamma^A = I,\; \gamma^{\mu},\;\gamma^{\mu}\gamma_{5},\;\sigma^{\mu\nu},\;\gamma_{5},
\]
apart from $\Gamma^A\! \equiv\! \sigma^{\mu\nu}$, the {\it vector} structure with $\Gamma^A\! \equiv \gamma^{\mu}$ is also different from zero for the case of Majorana spinors. This structure enters into the Lagrangian (\ref{eq:1y}) in the form $(1/e)\dot{x}_{\mu}(x)(\bar{\theta}\theta)(\bar{\psi}\gamma^{\mu} \psi)$. We shall analyze the mapping of this term. This mapping has a specific feature.
\\
\indent
Let us contract the initial expression (\ref{eq:2t}) with $\delta_{\beta \delta} (\gamma^{\mu})_{\gamma \alpha}$. Simple calculations result in
\[
\hbar\hspace{0.02cm}(\bar{\theta}\theta)(\bar{\psi}\hspace{0.03cm}\gamma^{\mu}\psi) = \frac{i}{4}\,\Bigl\{ -\bigl[\hspace{0.02cm}S\hspace{0.02cm}(V^{\mu})^{\ast} - V^{\mu}S^{\ast}\bigr] + \bigl[\hspace{0.02cm}P\hspace{0.02cm}(A^{\mu})^{\ast} - A^{\mu}P^{\ast}\bigr]
+ \bigl[\hspace{0.02cm}V_{\nu\hspace{0.02cm}}(T^{\mu\nu})^{\ast} - T^{\mu\nu}(V_{\nu})^{\ast}\bigr]
\]
\begin{equation}
+ \,\bigl[\hspace{0.02cm}A_{\nu}(\,^{\ast}T^{\mu\nu})^{\ast} - \,^{\ast}T^{\mu\nu}(A_{\nu})^{\ast}\bigr]\Bigr\},
\label{eq:2j}
\end{equation}
and, in particular, for the Majorana spinors we have
\begin{equation}
\hbar\hspace{0.02cm}(\bar{\theta}_{\rm M}\theta_{\rm M})(\bar{\psi}_{\rm M}\hspace{0.02cm}\gamma^{\mu}\hspace{0.02cm}\psi_{\rm M}) = -\frac{i}{2}\,\Bigl\{S\hspace{0.025cm}V^{\mu} - P\!\hspace{0.02cm}A^{\mu} - V_{\nu\,}T^{\mu\nu} - A_{\nu\!}\,^{\ast}T^{\mu\nu\!}\Bigr\}.
\label{eq:2k}
\end{equation}
Further, let us define a system of algebraic identities that have to satisfy the functions on the right-hand side of (\ref{eq:2k}). Here we need the equations of the ``vector'' type (\ref{ap:C2}) and (\ref{ap:C14}). For the Grassmann-valued currents they take the form
\begin{equation}
S\hspace{0.02cm}V^{\mu} = \frac{1}{2}\,P\!\hspace{0.02cm}A^{\mu} - \frac{1}{2}\,V_{\nu}\,T^{\mu\nu},\qquad
P\!\hspace{0.02cm}A^{\mu} = \frac{1}{2}\,S\hspace{0.02cm}V^{\mu} + \frac{1}{2}\,A_{\nu}\!\,^{\ast}T^{\mu\nu}
\label{eq:2l}
\end{equation}
or in a slightly different form they are
\begin{equation}
\begin{split}
3\bigl(&S\hspace{0.02cm}V^{\mu} - P\!\hspace{0.02cm}A^{\mu}\bigr) = - V_{\nu}\,T^{\mu\nu} - A_{\nu}\!\,^{\ast}T^{\mu\nu},\\
&S\hspace{0.02cm}V^{\mu} + P\!\hspace{0.02cm}A^{\mu} = - V_{\nu}\,T^{\mu\nu} + A_{\nu}\!\,^{\ast}T^{\mu\nu}.
\end{split}
\label{eq:2z}
\end{equation}
Moreover, one can obtain two more additional vector equations from  (\ref{ap:C7}) and (\ref{ap:C11}) contracting them with $\epsilon_{\mu\nu\lambda\sigma}$ and $g_{\lambda\nu}$, respectively. A somewhat cumbersome, but simple analysis shows that such obtained two additional equations are a consequence of (\ref{eq:2l}).\\
\indent
Equations (\ref{eq:2z}) enables us to eliminate a pair of variables $(SV^{\mu}, A_{\nu\!}\,^{\ast}T^{\mu \nu})$ in (\ref{eq:2k}):
\begin{equation}
\hbar\hspace{0.02cm}(\bar{\theta}_{\rm M}\theta_{\rm M})(\bar{\psi}_{\rm M}\gamma^{\mu}\psi_{\rm M}) = \hspace{0.02cm}i \bigl(P\!\hspace{0.02cm}A^{\mu} + V_{\nu}\,T^{\mu\nu}\bigr).
\label{eq:2zz}
\end{equation}
If we take into account (\ref{eq:2h}) and na\"{\i}vely set
\begin{equation}
P = \pm\,\xi_5
\label{eq:2x}
\end{equation}
in terms of the variables of the Lagrangian (\ref{ap:A1}), then for the first term on the right-hand side of the above expression we obtain rather unusual correspondence
\begin{equation}
\frac{1}{e}\,\dot{x}_{\mu}(x)(\bar{\theta}_{\rm M}\theta_{\rm M})(\bar{\psi}_{\rm M}\gamma^{\mu}\psi_{\rm M}) =
\frac{1}{\hbar}\,\frac{i}{e}\,\dot{x}_{\mu}\hspace{0.02cm}
\xi_{5\hspace{0.02cm}}\xi^{\mu}.
\label{eq:2c}
\end{equation}
In contrast to the tensor contribution (\ref{eq:2g}), the mapping of the term $(1/e)\dot{x}_{\mu}(x)(\bar{\theta}\theta)(\bar{\psi}\gamma^{\mu} \psi)$ results in the expression containing Planck's constant in the denominator. However, there is another problem. If one compares the expression obtained (\ref{eq:2c}) with the last term in (\ref{ap:A14}), namely with
\begin{equation}
\displaystyle\frac{i}{m\hspace{0.02cm}e}\,\dot{x}_{\mu}\hspace{0.02cm}\dot{\xi}_{5}\hspace{0.02cm}
\xi^{\mu},
\label{eq:2v}
\end{equation}
then it is seen that this expression is different from (\ref{eq:2c}) by the fact that instead of $\xi_5$ we have here $\dot{\xi}_5$. One can overcome these difficulties if instead of (\ref{eq:2x}) to use more nontrivial identification
\begin{equation}
P= \pm\, \frac{\hbar}{m}\,\dot{\xi}_5.
\label{eq:2b}
\end{equation}
By this means as in the case of the tensor contribution considered above, by discarding in (\ref{eq:2zz}) the terms which have no counterparts in expression (\ref{ap:A14}), one can achieve a good agreement with the reduced Lagrangian  (\ref{ap:A1}) (see the previous footnote).\\
\indent
In conclusion of this section we discuss briefly one more line investigation closely connected with the problem under consideration. In the early 1960s, the mathematician E. K\"ahler \cite{kahler_1962} has introduced a transcription of the Dirac equation as a set of equations for antisymmetric tensor fields (inhomogeneous differential forms). This approach received a further development in the papers \cite{graf_1978, benn_1983, becher_1982}. The geometrical description of spinor fields has been used extensively to formulate a consistent lattice formulation of fermions and in gravitation theory (it does not require the use of tetrad fields).\\
\indent
As was known, in the general case one K\"ahler fermion corresponds (in four dimensions) to four usual spinors. However, in the paper by G\"ursey \cite{gursey_1983} (see also \cite{jourjine_1987}) it was considered an important example when the K\"ahler fermion $\Psi$ can be constructed from {\it two} commuting (or anticommuting) Dirac spinors $(\psi, \varphi)$, namely the $4\times4$ matrix $\Psi$ consists of columns
\[
\bigl(\psi,\; \varphi,\; - \gamma_5 \psi^c,\; \gamma_5 \varphi^c\bigr),
\]
where $\psi^c$ and $\varphi^c$ are the charge-conjugate spinors. On the other hand, the matrix $\Psi$ can be expanded in elements generated by the matrices $\gamma_{\mu}$ as was made on the right-hand side of Eq.\,(\ref{eq:2q}). In principle, this gives an alternative relation to (\ref{eq:2q}) between a pair of two (anti)commuting spinors $(\psi,\, \varphi)$ and appropriate tensor set. As for our problem, the question arises whether one could extend the K\"ahler-G\"ursey approach to the case of a pair of spinors $(\psi,\, \theta)$ (or in the general case, four spinors) having different Grassmann parity? It would give a possibility of alternative construction of the required mapping.

%%%%%%%%%%%%%%%%%%%%%%%%%%%%%%%%%%%%%%%%%%%%%%%%%

\section{Mapping the kinetic term}
\setcounter{equation}{0}
\label{section_3}

Let us consider a mapping of the kinetic term in (\ref{eq:1y}), more exactly, of the term
\begin{equation}
\frac{1}{2}\,i\hspace{0.01cm}\hbar\hspace{0.05cm}(\bar{\theta}\theta)\!\left(\frac{d\bar{\psi}}{d\tau}\,
\psi - \bar{\psi}\,\frac{d\psi}{d\tau}\right).
\label{eq:3q}
\end{equation}
Here we have taken into account the choice of the constant $\lambda$, Eq.\,(\ref{eq:2ii}). Note also that the necessity of introducing the nilpotent factor for the construction of the correct mapping was first pointed out by Barut and Pav{\v s}i{\v c}  \cite{barut_pavsic_1989}. We shall assume for the moment that in the general case the auxiliary spinor $\theta_{\alpha}$ may be a function of $\tau$. Differentiating (\ref{eq:2w}) over $\tau$, we obtain
\[
\hbar^{1/2}\!\left(\frac{d\bar{\psi}_{\beta}}{d\tau}\,\theta_{\alpha} + \bar{\psi}_{\beta}\,\frac{d\theta_{\alpha}}{d\tau}\right) =
\]
\[
= \frac{1}{4}\,
\Bigl\{i\hspace{0.01cm}\dot{S}^{\hspace{0.02cm}\ast}\hspace{0.01cm}\delta_{\alpha\beta} + \dot{V}_{\mu}^{\ast}(\gamma^{\mu})_{\alpha\beta} -
\frac{i}{2}\,(\!\,^{\ast}\dot{T}_{\mu\nu})^{\ast}(\sigma^{\mu\nu}\gamma_{5})_{\alpha\beta} -
i\hspace{0.01cm}\dot{A}_{\mu}^{\ast}(\gamma^{\mu}\gamma_{5})_{\alpha\beta} - \dot{P}^{\hspace{0.02cm}\ast}(\gamma_{5})_{\alpha\beta}\Bigr\}.
\]
We contract the left-hand side of the expression with $\hbar^{1/2} \psi_{\beta} \bar{\theta}_{\alpha}$, whereas its right-hand side is contracted with the appropriate expression (\ref{eq:2q}). As a result, we have
\[
\hbar\!\left[(\bar{\theta}\theta)\!\left(\frac{d\bar{\psi}}{d\tau}\,\psi\right) - (\bar{\psi}\psi)\!\left(\bar{\theta}\,\frac{d\theta}{d\tau}\right)\right] =
- \frac{1}{4}\,
\Bigl\{\dot{S}^{\hspace{0.02cm}\ast\!}\hspace{0.03cm}S +
\dot{V}_{\mu}^{\ast}\hspace{0.02cm}V^{\mu} -
\frac{1}{2}\,(\!\,^{\ast}\dot{T}_{\mu\nu})^{\ast}\,^{\ast}T^{\mu\nu} - \hspace{0.01cm}\dot{A}_{\mu}^{\ast}\hspace{0.02cm}A^{\mu} - \dot{P}^{\hspace{0.02cm}\ast}\!\hspace{0.03cm}P\Bigr\}.
\]
Subtracting from the last expression its complex conjugation, we finally obtain
\begin{equation}
\hbar\hspace{0.04cm}(\bar{\theta}\theta)\!\left[\left(\frac{d\bar{\psi}}{d\tau}\,\psi\right) - \left(\bar{\psi}\,\frac{d\psi}{d\tau}\right)\right]  -
\hbar\hspace{0.04cm}(\bar{\psi}\psi)\!\left[\left(\frac{d\bar{\theta}}{d\tau}\,\theta\right) - \left(\bar{\theta}\,\frac{d\theta}{d\tau}\right)\right]  =
\label{eq:3w}
\end{equation}
\[
- \frac{1}{4}\,
\Bigl\{\!\bigl(\dot{S}^{\hspace{0.02cm}\ast\!}\hspace{0.03cm}S -
S^{\hspace{0.03cm}\ast\!}\hspace{0.02cm}\dot{S}\bigr) +
\bigl(\dot{V}_{\mu}^{\ast\hspace{0.03cm}}V^{\mu} -
V_{\mu}^{\ast\hspace{0.03cm}}\dot{V}^{\mu}\bigr)
- \frac{1}{2}\bigl((\,^{\ast}\dot{T}_{\mu\nu})^{\ast}\,^{\ast}T^{\mu\nu} - (\,^{\ast}T_{\mu\nu})^{\ast}\,^{\ast}\dot{T}^{\mu\nu}\bigr)
\]
\[
-\, \bigl(\dot{A}_{\mu}^{\ast}\hspace{0.03cm}A^{\mu} -
A_{\mu}^{\ast}\hspace{0.03cm}\dot{A}^{\mu}\bigr) -
\bigl(\dot{P}^{\hspace{0.02cm}\ast\!}\hspace{0.02cm}P -
P^{\hspace{0.02cm}\ast\!}\hspace{0.02cm}\dot{P}\bigr)\!\Bigr\}.
\]
This kinetic term is greatly simplified for Majorana spinors. Taking into account the fact that $(\bar{\psi}_{\rm M} \psi_{\rm M})=0$ and that the conditions (\ref{eq:2r}) hold, the general expression (\ref{eq:3w}) results in
\begin{equation}
\hbar\hspace{0.02cm}(\bar{\theta}_{\rm M}\theta_{\rm M})\!\!\left[\left(\!\frac{d\bar{\psi}_{\rm M}}{d\tau}\,\psi_{\rm M}\!\right)\! -
\!\left(\!\bar{\psi}_{\rm M}\,\frac{d\psi_{\rm M}}{d\tau}\!\right)\right]\! =
\label{eq:3e}
\end{equation}
\[
= \frac{1}{2}
\left\{S\,\frac{dS}{d\tau} + V_{\mu}\,\frac{d\hspace{0.02cm}V^{\mu}}{d\tau} - \frac{1}{2}\,^{\ast}T_{\mu\nu}\,\frac{d\,^{\ast}T^{\mu\nu}}{d\tau} - A_{\mu}\,\frac{dA^{\mu}}{d\tau}
- P\,\frac{dP}{d\tau}\right\}\!.
\]
We note that on the left-hand side the term with the derivative of $\theta_{\alpha}$ vanishes whether the auxiliary spinor is a function of $\tau$ or not. In Appendix D it is shown how this circumstance manifests on the right-hand side of (\ref{eq:3e}).\\
\indent
As in the case of the spin structure (\ref{eq:2o}) not all of the terms on the right-hand side of the expression (\ref{eq:3e}) are independent. The quadratic terms with the derivatives satisfy a certain algebraic system which is similar to the one (\ref{ap:C1})\,--\,(\ref{ap:C15}).  In particular, this system defines the relationships between the terms in (\ref{eq:3e}).\\
\indent
For obtaining the required system of relations we follow the same reasoning as was used in the previous section. Our first step is to multiply the expression (\ref{eq:2q}) by its derivative
\[
\hbar\hspace{0.02cm}(\bar{\theta}_{\beta}\psi_{\alpha})\hspace{0.02cm}\frac{d}{d\tau}
\,(\bar{\theta}_{\delta}\hspace{0.02cm}\psi_{\gamma}) = \frac{1}{16}\,
\Bigl\{-i\hspace{0.01cm}S\hspace{0.01cm}\delta_{\alpha\beta} + V_{\mu}(\gamma^{\mu})_{\alpha\beta} - \frac{i}{2}\,^{\ast}T_{\mu\nu}(\sigma^{\mu\nu}\gamma_{5})_{\alpha\beta} +
i\hspace{0.01cm}A_{\mu}(\gamma^{\mu}\gamma_{5})_{\alpha\beta} + P(\gamma_{5})_{\alpha\beta}\Bigr\}
\]
\begin{equation}
\times
\Bigl\{-i\hspace{0.005cm}\dot{S}\hspace{0.01cm}\delta_{\gamma\delta} + \dot{V}_{\lambda}(\gamma^{\lambda})_{\gamma\delta} -
\frac{i}{2}(\,^{\ast}\dot{T}_{\lambda\sigma})(\sigma^{\lambda\sigma}\gamma_{5})_{\gamma\delta} +
i\hspace{0.01cm}\dot{A}_{\lambda}(\gamma^{\lambda}\gamma_{5})_{\gamma\delta} + \dot{P}(\gamma_{5})_{\gamma\delta}\Bigr\}.
\label{eq:3r}
\end{equation}
On the right-hand side we will have every possible products of the functions $S, V_{\mu}, \!\,^{\ast}T_{\mu\nu},\ldots$ and their derivatives $\dot{S}, \dot{V}_{\mu}, \!\,^{\ast}\dot{T}_{\mu\nu},\,\ldots\,\,$. Let us consider in more detail the left-hand side of the above expression. For this purpose we contract the left-hand side of (\ref{eq:3r}) with the simplest spinor structure $\delta_{\beta\gamma}\delta_{\delta\alpha}$
\begin{equation}
\hbar\hspace{0.02cm}(\bar{\theta}_{\beta}\psi_{\alpha})\hspace{0.02cm}\frac{d}{d\tau}\hspace{0.03cm}
(\bar{\theta}_{\delta}\hspace{0.02cm}\psi_{\gamma})\hspace{0.02cm}\delta_{\beta\gamma}
\delta_{\delta\alpha} =
\hbar\hspace{0.02cm}\bigl[(\bar{\theta}\psi)(\dot{\bar{\theta}}\psi) + (\bar{\theta}\dot{\psi})(\bar{\theta}\psi)\bigr] =
-i\hspace{0.02cm}\hbar^{1/2}\bigl[S(\dot{\bar{\theta}}\psi) + (\bar{\theta}\dot{\psi})S\,\bigr].
\label{eq:3t}
\end{equation}
By virtue of the anticommutative character of the auxiliary spinor $\theta_{\alpha}$ and scalar function $S$, on the right-hand side of the last expression we have
\begin{equation}
i\hspace{0.02cm}\hbar^{1/2}\bigl[(\dot{\bar{\theta}}\psi) - (\bar{\theta}\dot{\psi})\bigr]S.
\label{eq:3y}
\end{equation}
Thus, the terms in square brackets do not collect into the required combination
\[
\hbar^{1/2}\bigl[(\dot{\bar{\theta}}\psi) + (\bar{\theta}\dot{\psi})\bigr] \equiv -i\hspace{0.02cm}\dot{S},
\]
thereby essentially complicate further analysis. For the sake of simplicity, in this section we restrict our consideration to the particular case
\[
\theta_{\alpha} = {\rm const}.
\]
In Appendix D we briefly discuss a way for overcoming the difficulty in the general case (at least for the Majorana spinors). \\
\indent
In this way, for the special case $\theta_{\alpha}\!\hspace{0.02cm} =\! \hspace{0.02cm}{\rm const.}$, from (\ref{eq:3t}) we have
\[
\hbar\hspace{0.02cm}(\bar{\theta}_{\beta}\psi_{\alpha})\,\frac{d}{d\tau}(\bar{\theta}_{\delta}
\hspace{0.02cm}\psi_{\gamma})\hspace{0.02cm}\delta_{\beta\gamma}\delta_{\delta\alpha} =
-\dot{S}S.
\]
Further, the contraction of the right-hand side of (\ref{eq:3r}) with $\delta_{\beta\gamma}\delta_{\delta\alpha}$ is readily calculated and we obtain the first desired algebraic equation containing the derivatives
\begin{equation}
4\dot{S}S =  S\dot{S} - P\dot{P} - V_{\mu}\dot{V}^{\mu} - A_{\mu}\dot{A}^{\mu} + \frac{1}{2}\hspace{0.01cm}\,^{\ast}T_{\mu\nu}\,^{\ast}\dot{T}^{\mu\nu}.
\label{eq:3u}
\end{equation}
The equation is a peculiar analog of Eq.\,(\ref{ap:C1}) with a fundamental distinction that  (\ref{ap:C1}) vanishes for the Grassmann-valued functions, whereas (\ref{eq:3u}) represents a rather nontrivial relation.\\
\indent
We specially have not collected similar terms with $S \dot{S}$ on the left- and right-hand sides to show how they arise in the original form from (\ref{eq:3r}). In particular, this enables us to understand how all remaining equations of a similar type can be directly obtained from the system  (\ref{ap:C2})\,--\, (\ref{ap:C15}) without recourse to the general formula (\ref{eq:3r}). For deriving the required equations it is sufficient on the left-hand side of each bilinear identity from  (\ref{ap:C1})\,--\, (\ref{ap:C15}) to make the replacement
\[
AB\rightarrow\dot{A}B,
\]
where $A$ and $B$ are any functions from the tensor set $(S, V_{\mu}, ^{\ast}\!T_{\mu \nu}, A_{\mu}, P)$ (i.e. the function with derivative must stand {\it on the left}), whereas on the right-hand side of identities  (\ref{ap:C1})\,--\, (\ref{ap:C15}) the function with derivative must stand {\it on the right}, as it takes place in (\ref{eq:3u}). Meanwhile we have to take into consideration the contributions both from commutators and from {\it anticommutators} by the following rules
\[
\begin{split}
&[\hspace{0.03cm}A,B\hspace{0.03cm}] = AB - BA \rightarrow  A\dot{B} - B\dot{A},\\
&\{A,B\} = AB + BA \rightarrow  A\dot{B} + B\dot{A}.
\end{split}
\]
Recall that the anticommutator is equal to zero for arbitrary Grassmann-odd functions $A$ and $B$ by definition. For the concrete expression (\ref{eq:3e}) in addition to equation (\ref{eq:3u}) we need the equations for $V_{\mu} \dot{V}^{\mu}, \,^{\ast}T_{\mu \nu}\,^{\ast}\dot{T}^{\mu \nu}, A_{\mu} \dot{A}^{\mu}$ and $P \dot{P}$. They follow from  (\ref{ap:C6}),  (\ref{ap:C13}), (\ref{ap:C15}) and  (\ref{ap:C10}) by the scheme given above
\begin{equation}
\begin{split}
&\dot{V}^{\mu}V_{\mu} =  - S\dot{S} - P\dot{P} - \frac{1}{2}\,V_{\mu}\dot{V}^{\mu} + \frac{1}{2}\,A_{\mu}\dot{A}^{\mu},\\
&\dot{A}^{\mu}A_{\mu} =  - S\dot{S} - P\dot{P} + \frac{1}{2}\,V_{\mu}\dot{V}^{\mu} - \frac{1}{2}\,A_{\mu}\dot{A}^{\mu},\\
&4\dot{P}P =  - S\dot{S} + P\dot{P} - V_{\mu}\dot{V}^{\mu} - A_{\mu}\dot{A}^{\mu} - \frac{1}{2}\hspace{0.01cm}\,^{\ast}T_{\mu\nu}\!\,^{\ast}\dot{T}^{\mu\nu},\\
&\,^{\ast}\dot{T}^{\mu\nu}\,^{\ast}T_{\mu\nu} = 3\hspace{0.02cm}(S\dot{S} - P\dot{P}) - \frac{1}{2}\hspace{0.01cm}\,^{\ast}T_{\mu\nu}\!\,^{\ast}\dot{T}^{\mu\nu}.
\end{split}
\label{eq:3i}
\end{equation}
Among the equations (\ref{eq:3u}) and (\ref{eq:3i}) only two ones are independent. It is convenient to write them in the form
\begin{equation}
2\hspace{0.01cm}(S\dot{S} + P\dot{P}) = V_{\mu}\dot{V}^{\mu} + A_{\mu}\dot{A}^{\mu},
\label{eq:3o}
\end{equation}
\begin{equation}
3\hspace{0.01cm}(S\dot{S} - P\dot{P}) = - \frac{1}{2}\hspace{0.01cm}\,^{\ast}T_{\mu\nu}\!\,^{\ast}\dot{T}^{\mu\nu}.
\label{eq:3p}
\end{equation}
These equations allow us to eliminate the terms $(-\frac{1}{2}) \,^{\ast}T_{\mu \nu}\,^{\ast}\dot{T}_{\mu\nu}$ and $V_{\mu} \dot{V}^{\mu}$ from the right-hand side of (\ref{eq:3e}). Multiplying (\ref{eq:3e}) by the factor $(i/2)$, we finally obtain
\begin{equation}
\frac{i}{2}\hspace{0.035cm}\hbar\hspace{0.02cm}(\bar{\theta}_{\rm M}\theta_{\rm M})\!\!\left[\left(\!\frac{d\bar{\psi}_{\rm M}}{d\tau}\,\psi_{\rm M}\!\right) -
\left(\!\bar{\psi}_{\rm M}\,\frac{d\psi_{\rm M}}{d\tau}\!\right)\right]\! =
-\frac{i}{2}\,A_{\mu}\dot{A}^{\mu} -\frac{i}{2}\,P\!\hspace{0.035cm}\dot{P} +
\frac{3\hspace{0.02cm}i}{2}\,S\dot{S}.
\label{eq:3a}
\end{equation}
We need to compare the right-hand side of this mapping with appropriate kinetic terms in the reduced expressions (\ref{ap:A14}) and (\ref{ap:A15}), namely with
\begin{equation}
-\hspace{0.015cm}\frac{i}{2}\,\xi_{\mu}\hspace{0.03cm}\dot{\xi}^{\mu} - \frac{i}{2}\,\xi_{5}\hspace{0.03cm}\dot{\xi}_{5}.
\label{eq:3s}
\end{equation}
The identification of the pseudovector $A_{\mu}$ with $\xi_{\mu}$ is exactly the same as that in the previous section, Eq.\,(\ref{eq:2h}). One needs to identify the pseudoscalar variable $P$ with $\xi_5$. At first sight the identification (\ref{eq:2x}) is the most natural and exactly reproduces the second term in (\ref{eq:3s}). However, such an identification contradicts (\ref{eq:2v}). From the other hand, in choosing (\ref{eq:2b}) we will have
\[
P\dot{P} = \frac{\hbar^2}{m^2}\,\dot{\xi}_{5}\hspace{0.03cm}\ddot{\xi}_5.
\]
In this case in order to obtain the correct expression (\ref{eq:3s}) we must require the fulfillment of the following equation for the variable $\xi_5$:
\begin{equation}
\ddot{\xi}_{5} = - \frac{m^2}{\hbar^2}\,\xi_5.
\label{eq:3d}
\end{equation}
This equation was first considered in the paper by Barut and Pa\v{v}si\v{c} \cite{barut_pavsic_1989}.\\
\indent
Further, we can eliminate the additional term with the $S$ function by setting $S=0$ (or in the more general case $S={\rm const.}$). Note that the condition $S=0$ coincides with one of the conditions of vanishing  the ``excess'' term in the tensor expression (\ref{eq:2f}) (see footnote \ref{foot_6}, though). By this means we see that in the case of Majorana spinors it is also possible to achieve the almost perfect mapping between the kinetic terms in the Lagrangians (\ref{eq:1t}) and (\ref{ap:A1}) (after eliminating the $\chi$-field) as in the case of mapping the force term.\\
\indent
At the close of Section 3 of the second part of our paper \cite{markov_part_II}, we shall discuss the possibility of identification of the $P$ and $\xi_5$ variables without using the constraint (\ref{ap:A13}) (see the remark immediately following Eq.\,(\ref{ap:A15})).

%
%%%%%%%%%%%%%%%%%%%%%% section 4 %%%%%%%%%%%%%%%%%%%%%
%

\section{\bf Mapping the bosonic symmetry}
\setcounter{equation}{0}
\label{section_4}

In this section we would like to discuss the symmetry transformations of the Lagrangian (\ref{eq:1t}) and to consider the problem of connection of this symmetry with the symmetry (\ref{ap:A16})\,--\,(\ref{ap:A20}) of the reduced Lagrangian (\ref{ap:A1}). The bosonic invariance for Lagrangians of the (\ref{eq:1t}) type was first discussed by Kowalski-Glikman {\it et al.} in \cite{kowalski-glikman_1988} in the free massless case and then in the paper by Barut and Pa\v{v}si\v{c} \cite{barut_pavsic_1989} with an extension to the massive particle. Let us consider the transformations in the form suggested in the latter paper (Eq.\,(39)) in the free case:
\begin{align}
&\delta{x}_{\mu} = \lambda\hspace{0.02cm}\bigl(\bar{\beta}\gamma_{\mu}\psi + \bar{\psi}\gamma_{\mu}\beta\bigr),  \label{eq:4q}\\
&\delta{e} = -\frac{2\hspace{0.015cm}i}{\hbar}\,\lambda \bigl(\bar{\beta}\hspace{0.015cm}\psi - \bar{\psi}\beta\bigr), \label{eq:4w}\\
&\delta\psi = -\frac{i}{\hbar}\,\frac{1}{e}\, \bigl[\hspace{0.02cm}\dot{x}_{\mu} - \lambda\hspace{0.02cm}(\bar{\psi}\gamma_{\mu}\psi)\bigr]\gamma^{\mu\!}\beta.  \label{eq:4e}
\end{align}
Hereafter, for brevity we reset again $\lambda \equiv (\bar{\theta} \theta)$. The function $\beta = \beta(\tau)$ is a {\it commuting} infinitesimal spinor parameter. For the time being we do not make any restrictions on the type of spinors $\psi, \theta$ and $\beta$ considering generally that they are Dirac ones. We shall demand the only condition $\dot{\theta} =0$. The free part of the Lagrangian (\ref{eq:1t}) transforms as follows
\begin{equation}
\delta L  \equiv \delta L_{0} + \delta L_{m}
 \label{eq:4r}
 \end{equation}
\begin{align}
= &-\lambda\,\frac{1}{e}\, \bigl[\hspace{0.02cm}\dot{x}_{\mu} - \lambda\hspace{0.02cm}(\bar{\psi}\hspace{0.02cm}\gamma_{\mu}\psi)\bigr] \frac{d}{d\tau}\, (\bar{\psi}\hspace{0.02cm}\gamma^{\mu}\beta) +
\lambda\,\frac{1}{2\hspace{0.015cm}e}\, \bigl[\hspace{0.02cm}\dot{x}_{\mu} - \lambda\hspace{0.02cm}(\bar{\psi}\hspace{0.02cm}\gamma_{\mu}\psi)\bigr] \bigl(\dot{\bar{\psi}}\hspace{0.02cm}\gamma^{\mu}\beta - \bar{\psi}\hspace{0.02cm}\gamma^{\mu\!}\dot{\beta}\bigr)   \notag \\
&- \lambda\hspace{0.02cm}(\bar{\psi}\gamma^{\mu}\beta)\, \frac{d}{d\tau}\biggl(\frac{1}{2\hspace{0.015cm}e}\,
\bigl[\hspace{0.02cm}\dot{x}_{\mu} \,-\, \lambda\hspace{0.02cm}(\bar{\psi}\gamma_{\mu}\psi)\bigr]\biggr) - \lambda\hspace{0.02cm}\frac{i\hspace{0.015cm}m^2}{\hbar}\,(\bar{\psi}\beta) \,+\, (\mbox{compl. conj.}). \notag
\end{align}
From an explicit form of the variation $\delta L$ we see that the transformations (\ref{eq:4q})\,--\,(\ref{eq:4e}) generally speaking, do not result in the desired invariance. Firstly, the first three terms on the right-hand side of (\ref{eq:4r}) do not collect in the total derivative. This is connected with the fact that in the second term the expression in parentheses has incorrect sign. Secondly, the last term in (\ref{eq:4r}) connected with the variation $\delta L_m$ remains uncompensated. To correct the situation, we consider a minimal modification of the transformations (\ref{eq:4q})\,--\,(\ref{eq:4e}), more exactly the modification of the transformation for the commuting spinor $\psi$:
\begin{equation}
\delta\psi = -\frac{i}{\hbar}\,\frac{1}{e}\, \bigl[\hspace{0.02cm}\dot{x}_{\mu} - \lambda\hspace{0.02cm}(\bar{\psi}\hspace{0.02cm}\gamma_{\mu}\psi)\bigr]\gamma^{\mu}\beta \,+\, \rho\hspace{0.015cm}\dot{\beta}.
\label{eq:4t}
\end{equation}
Here $\rho$ is some numerical parameter. The last term with $\dot{\beta}$ leads to appearance of two additional terms in the variation (\ref{eq:4r})
\[
\lambda\hspace{0.015cm}\rho\,\frac{1}{e}\, \bigl[\hspace{0.02cm}\dot{x}_{\mu} - \lambda\hspace{0.02cm}(\bar{\psi}\gamma_{\mu}\psi)\bigr]
\bigl(\bar{\psi}\gamma^{\mu}\dot{\beta}\bigr) \,+\,
\frac{1}{2}\,i\hbar\hspace{0.025cm}\lambda\rho\hspace{0.015cm}\biggl(\frac{d}{d\tau}
\hspace{0.03cm}
\bigl(\bar{\psi}\dot{\beta}\bigr) - 2 \bigl(\bar{\psi}\ddot{\beta}\hspace{0.015cm}\bigr)\!\biggr)  \,+\, (\mbox{compl. conj.}).
\]
Let us choose the parameter $\rho$ so that one could collect the total derivative from the first three terms in (\ref{eq:4r}). To do this, it is sufficient to set
\[
\rho = 1.
\]
To eliminate the last term in (\ref{eq:4r}) some restriction on the $\tau\hspace{0.025cm}$-\,dependence of the spinor $\beta$ is required, namely, we demand the fulfillment of the following equation:
\begin{equation}
\ddot{\beta}\, + \frac{m^2}{\hbar^2}\,\beta = 0.
\label{eq:4y}
\end{equation}
It is precisely this condition for the pseudoscalar variable $\xi_5$ that has arisen at the end of the previous section, Eq.\,(\ref{eq:3d}).\\
\indent
Now we turn to the problem of a connection of symmetry transformations (\ref{eq:4q}), (\ref{eq:4w}) and (\ref{eq:4t}) with the reduced transformations of local supersymmetry (\ref{ap:A16})\,--\,(\ref{ap:A19}). Here we restrict our attention to Majorana spinors only (in what follows, for brevity, we omit the symbol $M$ for the Majorana spinors). Let us consider the mapping of transformation (\ref{eq:4q}). Taking into account that for Majorana spinors the equality $\bar{\beta} \gamma_{\mu} \psi = \bar{\psi} \gamma_{\mu}\hspace{0.015cm}\beta$ holds, we have
\[
\delta{x}_{\mu} = 2\hspace{0.015cm}\lambda\hspace{0.02cm}(\bar{\psi}\gamma_{\mu}\beta) \equiv
2\hspace{0.015cm}(\bar{\theta}\theta)(\bar{\psi}\gamma_{\mu}\beta).
\]
From the other hand, by virtue of (A.16), the definitions (\ref{eq:2e}) and identification (\ref{eq:2h}), we can write
\[
\delta{x}_{\mu}=i\hspace{0.02cm}\alpha\hspace{0.02cm}\xi_{\mu} \equiv \pm\hspace{0.02cm} i\hspace{0.02cm}\alpha\hspace{0.015cm}A_{\mu} =
\pm\hspace{0.02cm}\hbar^{1/2}\alpha\hspace{0.02cm}(\bar{\psi}\gamma_{\mu}\gamma_{5}\theta).
\]
Comparing these two expressions, we obtain a connection between the commuting spinor  $\beta$ of bosonic transformations and the Grassmann scalar parameter $\alpha$ of supertransformations
\begin{equation}
(\bar{\theta}\theta)\beta = \pm\hspace{0.015cm}\frac{1}{2}\,\hbar^{1/2}\alpha\hspace{0.02cm}(\gamma_{5}\theta).
\label{eq:4u}
\end{equation}
For the constant auxiliary spinor $\theta$ equation (\ref{eq:4y}) for the spinor parameter  $\beta$ turns to a similar equation for the scalar parameter $\alpha$.\\
\indent
Let us consider further the mapping of transformation of the einbein field $e$,  Eq.\,(\ref{eq:4w}). With allowance for (\ref{eq:4u}), the property $(\bar{\beta}\hspace{0.015cm}\psi) = -\hspace{0.015cm} (\bar{\psi} \beta)$ for Majorana spinors and the definitions (\ref{eq:2e}), we have here the chain of equalities
\[
\delta{e} = \frac{4\hspace{0.015cm}i}{\hbar}\,\lambda\hspace{0.015cm}(\bar{\psi}\beta) =
\pm\hspace{0.015cm}\frac{2\hspace{0.015cm}i}{\hbar^{1/2}}\,
\alpha\hspace{0.015cm}(\bar{\psi}\gamma_{5}\theta)
\equiv
\mp\hspace{0.015cm}\frac{2\hspace{0.015cm}i}{\hbar}\,
\alpha P.
\]
From the other hand, for the reduced transformation of supersymmetry, Eq.\,(\ref{ap:A18}), we have
\[
\delta e = - \frac{2\hspace{0.015cm}i}{m}\,\alpha\,\dot{\xi}_5.
\]
Comparing these two expressions we obtain that
\begin{equation}
P = \pm\hspace{0.015cm}\frac{\hbar}{m}\;\dot{\xi}_5.
\label{eq:4i}
\end{equation}
It coincides in exact with our choice for the representation of the function $P$, Eq.\,(\ref{eq:2b}).\\
\indent
We proceed now to analysis of mapping the transformation for the commuting spinor $\psi$, Eq.\,(\ref{eq:4t}). First we multiply the expression (\ref{eq:4t}) from the left by $i\lambda\hbar^{1/2}\bar{\theta}\gamma_{\mu}\gamma_5$. Taking into account the connection (\ref{eq:4u}) and the identity $\gamma_{\mu}\gamma_{\nu} = I \!\cdot\! g_{\mu \nu} + i \sigma_{\mu \nu}$, we get
\begin{equation}
i\hspace{0.015cm}\lambda\hspace{0.015cm}\hbar^{1/2}\delta(\bar{\theta}\gamma_{\mu}\gamma_5\psi)
=
 \label{eq:4o}
\end{equation}
\[
= \mp\,\frac{1}{e}\,\lambda\hspace{0.015cm}\alpha\bigl[\hspace{0.02cm}\dot{x}_{\mu} - (\bar{\theta}\theta)\hspace{0.02cm}(\bar{\psi}\gamma_{\mu}\psi)\bigr]
\mp\frac{i}{e}\,\alpha\hspace{0.02cm}(\bar{\theta}\sigma_{\mu \nu}\theta)
\bigl[\hspace{0.02cm}\dot{x}^{\nu} - \lambda\hspace{0.02cm}(\bar{\psi}\gamma^{\nu}\psi)\bigr]
\hspace{0.02cm}\pm\hspace{0.02cm}i\hspace{0.015cm}\hbar\hspace{0.03cm}\dot{\alpha}
\hspace{0.015cm}(\bar{\theta}\gamma_{\mu}\theta).
\]
In deriving the above expression we have considered that the parameter $\alpha$ and the spinor
$\theta_{\alpha}$ commute with one another. For Majorana spinors the last two terms become zero by virtue of the properties (\ref{eq:2u}). Further we use the exact relation (\ref{eq:2zz}), where for the function $P$ we take (\ref{eq:4i}). Dropping the contribution with $V_{\nu\,}T^{\mu\nu}$ in (\ref{eq:2zz}) and taking into account that
\[
A_{\mu} \equiv i\hspace{0.015cm}\hbar^{1/2}(\bar{\theta}\gamma_{\mu}\gamma_{5\hspace{0.02cm}}\psi) =
\pm\hspace{0.02cm}\xi_{\mu},
\]
we finally obtain
\begin{equation}
\lambda\hspace{0.025cm}\delta{\xi}_{\mu} = -\lambda\hspace{0.02cm}\alpha\Bigl(\dot{x}_{\mu} -
\displaystyle\frac{i}{m}\,\dot{\xi}_{5}\hspace{0.02cm}\xi_{\mu}\Bigr)\!\Big/e.
 \label{eq:4p}
\end{equation}
Thus, by simple dropping excess terms in the mapping (\ref{eq:2zz}), one can achieve exact coincidence with the reduced transformation of supersymmetry (\ref{ap:A17}) (more exactly,  within the overall nilpotent factor $\lambda \equiv (\bar{\theta} \theta)$).\\
\indent
It remains to reproduce the transformation (\ref{ap:A19}). For this purpose we multiply the transformation (\ref{eq:4t}) from the left by $\lambda\hbar^{1/2}\bar{\theta}\gamma_5$:
\[
\lambda\hspace{0.015cm}\hbar^{1/2}\delta(\bar{\theta}\gamma_5\psi) =
-\hspace{0.015cm}i\hspace{0.015cm}\lambda\,\frac{1}{\hbar^{1/2}}\,\frac{1}{e}\,\bigl[\hspace{0.02cm}
\dot{x}_{\mu} - \lambda\hspace{0.02cm}(\bar{\psi}\gamma_{\mu}\psi)\bigr]
(\bar{\theta}\gamma_{5}\gamma^{\mu}\beta)
\,+\lambda\hspace{0.01cm}\hbar^{1/2}(\bar{\theta}\gamma_{5}\dot{\beta}).
\]
Taking into account the relation for the spinor $\beta$, Eq.\,(\ref{eq:4u}) and the remark following Eq.\,(\ref{eq:4o}), we get
\begin{equation}
\lambda\hspace{0.015cm}\hbar^{1/2}\delta(\bar{\theta}\gamma_5\psi) =
\pm\hspace{0.03cm} i\hspace{0.015cm}\alpha\,\frac{1}{2\hspace{0.015cm}e}\,(\bar{\theta}\gamma^{\mu}\theta)
\bigl[\hspace{0.02cm}\dot{x}_{\mu} - \lambda\hspace{0.02cm}(\bar{\psi}\gamma_{\mu}\psi)\bigr]
\hspace{0.015cm}\pm\hspace{0.015cm}\frac{1}{2}\,\lambda\hspace{0.015cm}\dot{\alpha}\hbar.
\label{eq:4a}
\end{equation}
For Majorana spinors the first term on the right hand side is zero. Recalling the definition of the function $P \equiv \hbar^{1/2} (\bar{\theta} \gamma_5 \psi)$ and identification
(\ref{eq:4i}), we finally obtain
\[
\lambda\hspace{0.015cm}\delta\dot{\xi}_5 = \lambda\hspace{0.015cm} m \dot{\alpha}.
\]
Such approach enables us to reproduce not the transformation (A.19) itself, but its derivative (up to the overall nilpotent factor $\lambda$, as in the previous case). We specially note that in the mapping (\ref{eq:4a}) only the term connected with the last one in the modified transformation (\ref{eq:4t}), survives. This is used for additional confirmation of the necessity of the presence of the term with $\dot{\beta}$ in (\ref{eq:4t}) for correct reproduction of supersymmetry transformation.\\
\indent
Next it would be necessary to consider the transformations of bosonic symmetry and its mapping for the model with the interaction, i.e. with allowance made for the Lagrangian (\ref{eq:1i}). Here, we restrict our consideration to a few remarks of the general character.\\
\indent
The transformation for the Grassmann-valued color charge $\theta^i$
\[
\delta{\theta}^{i} =  -\frac{i}{\hbar}\,\lambda\hspace{0.015cm} g\hspace{0.02cm}\bigl(\bar{\beta}\gamma^{\mu}\psi + \bar{\psi}\gamma^{\mu}\beta\bigr)A_{\mu}^{a}(x)(t^{a})^{ij}\theta^{j}
\]
needs to be added to the symmetry transformations (\ref{eq:4q}), (\ref{eq:4w}) and (\ref{eq:4t}).
However, it is not enough to provide the invariance of Lagrangian (\ref{eq:1t}). The additional contributions to the symmetry transformations considered above, involving the background Yang-Mills field are required. For example, the transformation for the commuting spinor $\psi$, Eq.\,(\ref{eq:4t}), should be replaced now by
\begin{equation}
\delta\psi = -\frac{i}{\hbar}\,\frac{1}{e}\,\bigl[\hspace{0.02cm}\dot{x}_{\mu} - \lambda\hspace{0.02cm}(\bar{\psi}\gamma_{\mu}\psi)\bigr]\gamma^{\mu\!}\beta \,+ \dot{\beta}\,
- \frac{i\hspace{0.01cm}e\hspace{0.015cm}g}{4}\,Q^{a\!} F^{a}_{\mu\nu\,}\sigma^{\mu\nu\!}\beta.
 \label{eq:4s}
\end{equation}
These additional contributions to the symmetry transformations have to provide not only the invariance of the complete Lagrangian (\ref{eq:1t}). They must vanish under the mapping into the supertransformations (\ref{ap:A16})\,--\,(\ref{ap:A20}). Here, it is required a more accurate consideration of what one should understand as the mapping of the transformations. For instance, in the case of the mapping of transformation (\ref{eq:4s}), in the form of Eq.\,(\ref{eq:4o}), the additional term
\begin{equation}
\alpha\,\frac{e\hspace{0.015cm}g}{8}\,Q^{a\!} F^{a\hspace{0.015cm}\nu\lambda}(\bar{\theta}\gamma_{\mu}\sigma_{\nu\lambda}\theta),
 \label{eq:4d}
\end{equation}
which is not zero even for the Majorana spinors, appears. However, if on the left hand side of (\ref{eq:4o}) one uses a more accurate expression
\[
\frac{1}{2}\,i\hspace{0.015cm}\lambda\hspace{0.015cm}\hbar^{1/2}\bigl[
(\bar{\theta}\gamma_{\mu}\gamma_5\delta\psi) - (\delta\bar{\psi}\gamma_{\mu}\gamma_5\theta)\bigr],
\]
then, instead of (\ref{eq:4d}), we will already have
\[
\alpha\,\frac{e\hspace{0.015cm}g}{16}\,Q^{a\!} F^{a\hspace{0.015cm}\nu\lambda}
(\bar{\theta}\hspace{0.015cm}[\hspace{0.015cm}\gamma_{\mu},\sigma_{\nu\lambda}]
\hspace{0.015cm}\theta).
\]
Taking into account that by virtue of formulae (\ref{ap:B2}) the following relation holds
\[
[\hspace{0.015cm}\gamma_{\mu},\sigma_{\nu\lambda}] = \frac{2}{i}\,
(g_{\mu\nu}\gamma_{\lambda} - g_{\mu\lambda}\gamma_{\nu}),
\]
and that for the Majorana anticommuting spinors the equality $(\bar{\theta} \gamma_{\mu} \theta) = 0$ is true, we obtain that really under the mapping of transformation (\ref{eq:4s}) the last term in (\ref{eq:4s}) does not give a contribution and ipso facto we return again to the expression (\ref{eq:4p}).

%%%%%%%%%%%%%%%%%%%%%%%%%%%%%%%%%%%%%%%%%%%%%%%%%

\section{\bf Mapping into Lagrangian with a local supersymmetry}
\setcounter{equation}{0}
\label{section_9}

In the preceding sections it was considered the mapping of the original Lagrangian (\ref{eq:1t}) possessing (local) bosonic invariance into the Lagrangian obtained after the elimination of the variable $\chi$ from the Lagrangian (\ref{ap:A1}). The terms containing the fermion counterpart $\chi$ to the einbein field $e$, namely
\begin{equation}
\displaystyle\frac{i}{2\hspace{0.015cm}e}\,\chi\hspace{0.02cm}\dot{x}_{\mu}\hspace{0.02cm}\xi^{\mu},
\quad
\displaystyle\frac{im}{2}\,\chi\hspace{0.02cm}\xi_{5},
\label{eq:9q}
\end{equation}
cannot appear in principle under any map because there are no counterparts for them in the Lagrangian (\ref{eq:1t}). These terms are important for local supersymmetry of the Lagrangian (\ref{ap:A1}), and its counterparts a priory must be contained in the initial Lagrangian (\ref{eq:1t}). In this section we would like to show how terms of this kind may really appear.\\
\indent
The basic idea in determining such terms is that of using an extended Hamiltonian or superHamiltonian for the construction of the ``spinning'' equation (\ref{eq:1q}). Hamiltonian of this type has been considered in a few papers for various reasons. Thus in the papers by Di Vecchia and Ravndal \cite{vecchia_1979}, Ravndal \cite{ravndal_1980}, Borisov and Kulish \cite{borisov_1982}, Fradkin and Gitman \cite{fradkin_1991}, van Holten \cite{holten_1995} it has been used in the construction of the path integral representation for the Green's function of a Dirac particle in a background gauge field. Within the framework of operator formalism this superHamiltonian in the non-Abelian case has the form
\begin{equation}
-2\hspace{0.03cm}m\hat{H}_{\rm SUSY} =\Bigl( -\hbar^{2\!}\hspace{0.02cm}\hat{D}_{\mu}\hat{D}^{\mu}
+  \frac{g\hbar}{2}\,\hat{\sigma}^{\mu\nu\!}F^{a}_{\mu\nu\,}\hat{t}^a  - m^2 \Bigr)
\!+\hspace{0.02cm}i\hspace{0.01cm}m\hspace{0.015cm}\hbar^{1/2\!}\hspace{0.015cm}
\chi\bigl(\hspace{0.01cm}i\hbar\,\hat{\gamma}_{\mu}\hat{D}^{\mu} - m \bigr).
\label{eq:9w}
\end{equation}
All quantities with hats above represent operators acting in appropriate spaces of representa\-tions of the spinor, color and coordinate algebras; $\chi$ is an odd variable. The factor $m\hspace{0.015cm}\hbar^{1/2}$ in front of the $\chi$ is chosen thus that this function has the dimension coinciding with that of the one-dimensional gravitino field $\chi$ in the terms (\ref{eq:9q}). Analog of introducing such a superHamiltonian in the massless limit can be also found in the work of Friedan and Windey \cite{friedan_1984}. The superHamiltonian was used in the construction of superheat kernel. The latter has been used in calculating the chiral anomaly. In the monograph by Thaller \cite{thaller_book} within the supersymmetric quantum mechanics a notion of the {\it Dirac operator with supersymmetry} has been defined in the most general abstract form. The expression (\ref{eq:9w}) is just its special case.\\
\indent
Before studying the general case of the Dirac operator with supersymmetry it is necessary to recall briefly the fundamental points of deriving the equation of motion for the commuting spinor $\psi_{\alpha}$, Eq.\,(\ref{eq:1q}). The equation arises when we analyze the connection of the relativistic quantum mechanics with the relativistic classical mechanics first performed by Pauli \cite{pauli_1932} within the so-called first-order formalism for fermions (see, also \cite{bolte_1999}). In the paper by Fock \cite{fock_1937} and in the book by Akhiezer and Beresteskii \cite{akhiezer_1969} this analysis was performed on the basis of the second-order formalism. Here, we will follow the latter line.\\
\indent
In the second-order formalism the original QCD Dirac equation for a wave function $\Psi$ is replaced by its quadratic form
\begin{equation}
-2\hspace{0.03cm}m\hspace{0.02cm}\hat{H}\Phi =
\Bigl(-\hbar^{2\!}\hspace{0.02cm}{D}_{\mu}{D}^{\mu}
+  \frac{g\hbar}{2}\,{\sigma}^{\mu\nu\!}F^{a}_{\mu\nu\,}{t}^a  - m^2 \Bigr)\Phi = 0,
\label{eq:9e}
\end{equation}
where  ${D}_{\mu}(x) = {\rm I}\cdot \partial_{\mu} + i\hspace{0.02cm}(g/\hbar)\hspace{0.01cm} A^{a}_{\mu}(x)t^{a}$, I is the identity color matrix, and a new spinor $\Phi$ is connected with the initial one by the relation
\[
\Psi = \frac{1}{m}\,\Bigl(\hspace{0.01cm}i\hbar\,{\gamma}_{\mu}{D}^{\mu} + m \Bigr)\Phi.
\]
Since we are interested in the interaction of the spin degrees of freedom of a particle with an external gauge field most, then for the sake of simplicity we will consider equation (\ref{eq:9e}) for the case of the interaction with an Abelian background field (with the replacement of the coupling $g$ by $q$). The presence of the color degrees of freedom can result in qualitatively new features, one of them is appearing a mixed spin-color degrees of freedom \cite{arodz_1_1982}. In this respect our original model Lagrangian (\ref{eq:1t}) is the simplified one. It corresponds to perfect factorization of the spin and color degrees of freedom of the particle. The non-Abelian case also requires appreciable complication of the usual WKB-method in the analysis of Eq.\,(\ref{eq:9e}) that is beyond the scope of this work (see, for example, \cite{arodz_2_1982, belov_1992, guhr_2007}).\\
\indent
A solution of equation (\ref{eq:9e}) in the semiclassical limit is defined as a series in powers of $\hbar$
\begin{equation}
\Phi = {\rm e}^{iS/\hbar} (f_{0} + \hbar f_{1} + \hbar^{2\!}f_{2}\,+\,\ldots\,),
\label{eq:9r}
\end{equation}
where $S,\,f_0,\,f_1,\,\ldots$ are some functions of coordinates and of time. Substituting this series into (\ref{eq:9e}) and collecting terms of the same power in $\hbar$, we obtain the following equations correct to the first order in $\hbar$
\begin{equation}
\hbar^{0}:\hspace{3cm}
\biggl(\frac{\partial S}{\partial x^{\mu}} \,+\, q A_{\mu}\biggr)^{\!\!2}\! - m^{2} = 0,\hspace{5cm}
\label{eq:9t}
\end{equation}
\begin{equation}
\hbar^{1}:
\biggl[\,\frac{1}{i}\hspace{0.03cm}\frac{\partial}{\partial x_{\mu}}
\biggl(\frac{\partial S}{\partial x^{\mu}} \,+\, q A_{\mu}\biggr)\biggr]\hspace{0.02cm}f_{0} \,+\,
\frac{2}{i}\hspace{0.02cm}\biggl(\frac{\partial S}{\partial x^{\mu}} \,+\, q A_{\mu}\biggr)
\frac{\partial f_{0}}{\partial x_{\mu}} \,+\,
\frac{q}{2}\,\sigma_{\mu\nu}F^{\mu\nu}\! f_{0} =\hspace{0.02cm} 0.
\label{eq:9y}
\end{equation}
Further, we introduce into consideration the flux fermion density
\begin{equation}
s_{\mu} \equiv \bar{\Psi}_{0}\gamma_{\mu}\Psi_{0},
\label{eq:9u}
\end{equation}
where as $\Psi_0$ we take the following expression:
\[
\begin{split}
&\Psi_{0} = \frac{1}{m}\,\bigl(i\hbar\hspace{0.03cm}\gamma_{\mu}D^{\mu} + m \bigr)
f_{0\,}{\rm e}^{iS/\hbar}
\simeq
-\frac{1}{m}\,{\rm e}^{iS/\hbar}\bigl[\pi_{\mu}\gamma^{\mu} -
m \bigr]\hspace{0.01cm}f_{0},\\
&\pi_{\mu}\equiv\frac{\partial S(x,\boldsymbol{\alpha})}{\partial x^{\mu}} \,+\, q A_{\mu}(x).
\end{split}
\]
Here, $\boldsymbol{\alpha}$ designates three arbitrary constants defining a solution for the action $S$, Eq.\,(\ref{eq:9t}). In terms of the spinor $f_{0}$ the flux density (\ref{eq:9u}) takes the form
\[
s_{\mu} = \frac{2\,}{m^2}\,\pi_{\mu}\bigl[\hspace{0.03cm}\bar{f}_{0}(\gamma_{\nu}\pi^{\nu} - m)f_{0}\bigr]
\]
and, by virtue of Eqs.\,(\ref{eq:9t}) and (\ref{eq:9y}), it satisfies the equation of continuity
\[
\frac{\partial s_{\mu}}{\partial x_{\mu}} = 0.
\]
\indent
Equation (\ref{eq:1q}) arises from an analysis of the Eq.\,(\ref{eq:9y}) for the spinor $f_{0}$. Eq.\,(\ref{eq:9y}) in terms of the function $\pi_{\mu}$ can be written in a more compact form
\begin{equation}
\frac{\partial\hspace{0.02cm} \pi_{\mu}}{\partial x_{\mu}}\,f_{0} \,+\, 2\hspace{0.02cm}\pi_{\mu}\,\frac{\partial f_{0}}{\partial x_{\mu}} \,+ \,
\frac{i\hspace{0.02cm}q}{2}\,\sigma_{\mu\nu}F^{\mu\nu}\! f_{0} =\hspace{0.02cm} 0.
\label{eq:9i}
\end{equation}
At this point we introduce a new variable
\[
\eta\equiv \frac{2\,}{m^2} \bigl[\hspace{0.03cm}\bar{f}_{0}(\gamma_{\nu}\pi^{\nu} - m)f_{0}\bigr],
\]
such that $s_{\mu} = \pi_{\mu} \eta$. Owing to the continuity equation we have an important relation for the $\eta$ function
\begin{equation}
\frac{\partial\hspace{0.02cm} \pi_{\mu}}{\partial x_{\mu}}\,\eta = -\pi_{\mu}\,\frac{\partial\hspace{0.02cm}\eta}{\partial x_{\mu}}.
\label{eq:9o}
\end{equation}
At the final stage we substitute $f_{0}=\sqrt{\eta}\hspace{0.04cm}\varphi_{0}$ into Eq.\,(\ref{eq:9i}). In terms of a new spinor function $\varphi_{0}$ with allowance for (\ref{eq:9o}) this equation takes the following form:
\[
\pi_{\mu}\hspace{0.03cm}\frac{\partial\hspace{0.02cm} \varphi_{0}}{\partial x_{\mu}} = -\frac{i\hspace{0.02cm}q}{4}\,\sigma_{\mu\nu}F^{\mu\nu}\! \varphi_{0}.
\]
In the book \cite{akhiezer_1969} a solution of the equation obtained just above, was expressed in terms of the solution of Schr\"odinger's equation for the wave function $\psi_{\alpha}(\tau)$,
Eq.\,(\ref{eq:1q}). The latter describes the motion of a spin in a given gauge field $F_{\mu \nu}(x)$. This field is defined along the trajectory of the particle
$x_{\mu} = x_{\mu}(\tau,\boldsymbol{\alpha},\boldsymbol{\beta})$, which in turn is defined from a solution of the equation
\[
m\,\frac{d\hspace{0.02cm} x_{\mu}}{d\hspace{0.02cm}\tau\,} = \pi_{\mu}(x,\boldsymbol{\alpha})
\]
with the initial value given by a vector $\boldsymbol{\beta}$.\\
\indent
Let us now discuss the question of a modification of the above equations in the case when instead of the usual Hamilton operator in equation (\ref{eq:9e}) one takes its supersymmetric extension, i.e., considers the equation in the form
\begin{equation}
-2\hspace{0.025cm}m\hspace{0.02cm}\hat{H}_{\rm SUSY}\Phi
\equiv
\Bigl\{\!\Bigl(-\hbar^{2\!}\hspace{0.03cm}D_{\mu}D^{\mu} + \frac{q\hbar}{2}\,\sigma_{\mu\nu}F^{\mu\nu} - m^2\Bigr)
+\hspace{0.02cm}
i\hspace{0.01cm}m\hspace{0.015cm}\hbar^{1/2\!}\hspace{0.015cm}
\chi\bigl(i\hbar\,\gamma_{5}\gamma_{\mu}D^{\mu} -\, m\hspace{0.02cm}\gamma_{5}\bigr)\!\Bigr\}\,\Phi = 0.
\label{eq:9p}
\end{equation}
Here,  following \cite{fradkin_1991} in second parentheses we have introduced the $\gamma_5$ matrix into the definition of the linear Dirac operator. This operator $\bigl(i\hbar\,\gamma_{5}\gamma_{\mu\!}D^{\mu} - m \gamma_5\bigr)$ should be believed as an odd function. The expansion (\ref{eq:9r}) is modified as follows
\begin{equation}
\Phi = {\rm e}^{iS/\hbar} (f_{0} + \hbar^{1/2}\hspace{0.015cm}\chi\hspace{0.015cm} f_{\frac{1}{2}} + \hbar \hspace{0.015cm}f_{1} + \hbar^{\,3/2\!}\hspace{0.02cm}\chi\hspace{0.015cm} f_{\frac{3}{2}}\,+\,\ldots\,).
\label{eq:9a}
\end{equation}
In the decomposition (\ref{eq:9a}) we believe the functions $(S, f_{0}, f_{1}, \dots)$ to be commuting ones, and $(f_{\frac{1}{2}}, f_{\frac{3}{2}}, \ldots)$ to be anticommuting ones.
%The opposite case of the partition into the Grassmann parity will be mentioned at the end of this section.
Instead of Eqs.\, (\ref{eq:9t}) and (\ref{eq:9y}) we now have
\[
\begin{split}
%
%%%%%%%%%%%%%%%%%%%%% 0 %%%%%%%%%%%%%%%%%%%%%%%%%%
%
\hbar^{0\;\;\;\,}:\hspace{0.5cm}   &\bigl(\hspace{0.03cm}\pi^2 - m^2\bigr)f_{0} = 0,\\
%
%%%%%%%%%%%%%%%%%%%%% 1/2 %%%%%%%%%%%%%%%%%%%%%%%%%
%
\hbar^{1/2}:\hspace{0.5cm}   &\bigl(\hspace{0.03cm}\pi^2 - m^2\bigr)f_{\frac{1}{2}} -\, i\hspace{0.02cm}m \bigl(\hspace{0.03cm}\pi_{\mu}\gamma^{\mu}\gamma_{5} +\, m\hspace{0.02cm}\gamma_{5}\bigr)f_{0} = 0,\\
%
%%%%%%%%%%%%%%%%%%%%% 1 %%%%%%%%%%%%%%%%%%%%%%%%%%
%
\hbar^{1\;\;\;\,}:\hspace{0.5cm}   &\bigl(\hspace{0.03cm}\pi^2 - m^2\bigr)f_{1}
\,+\,
\biggl[\hspace{0.02cm}\frac{1}{i}\,\frac{\partial\hspace{0.02cm}
\pi_{\mu}}{\partial x_{\mu}}\,f_{0} \,+\, \frac{2}{i}\,\pi_{\mu}\,\frac{\partial f_{0}}{\partial x_{\mu}}
\,+\,
\frac{q}{2}\,\sigma_{\mu\nu}F^{\mu\nu}\! f_{0}\biggr] = 0,\\
%
%%%%%%%%%%%%%%%%%%%%% 3/2 %%%%%%%%%%%%%%%%%%%%%%%%%
%
\hbar^{3/2}:\hspace{0.5cm}   &\bigl(\hspace{0.03cm}\pi^2 - m^2\bigr)f_{\frac{3}{2}}
-\, i\hspace{0.02cm}m \bigl(\hspace{0.03cm}\pi_{\mu}\gamma^{\mu}\gamma_{5} +\, m\hspace{0.02cm}\gamma_{5}\bigr)f_{1}\; +\\
&\hspace{2.35cm}+
\biggl[\hspace{0.02cm}\frac{1}{i}\,\frac{\partial\hspace{0.02cm} \pi_{\mu}}
{\partial x_{\mu}}\,f_{\frac{1}{2}} \,+\, \frac{2}{i}\,\pi_{\mu}\,\frac{\partial f_{\frac{1}{2}}}
{\partial x_{\mu}}
\,+\,
\frac{q}{2}\,\sigma_{\mu\nu}F^{\mu\nu}\! f_{\frac{1}{2}}\biggr]
-\,
m\hspace{0.02cm}\gamma_{\mu}\gamma_{5}\,\frac{\partial f_{0}}{\partial x_{\mu}} = 0.
\end{split}
\]
The first equation defines the Hamilton-Jacobi equation for the action $S$, Eq.\,(\ref{eq:9t}), and the second one is reduced to the matrix algebraic equation for the spinor $f_{0}$
\[
\bigl(\hspace{0.03cm}\pi_{\mu}\gamma^{\mu}\gamma_{5} +\, m\hspace{0.02cm}\gamma_{5}\bigr)f_{0} = \hspace{0.02cm} 0.
\]
Further, the equation of first order in $\hbar$ reproduces Eq.\,(\ref{eq:9i}). Finally, we will rewritten the last equation in a somewhat more convenient form for the subsequent analysis
\begin{equation}
\frac{1}{i}\biggl(\frac{\partial\hspace{0.02cm} \pi_{\mu}}{\partial x_{\mu}}\biggr)\!f_{\frac{1}{2}} \,+\,
 \frac{2}{i}\,\pi_{\mu}\hspace{0.02cm}\frac{\partial f_{\frac{1}{2}}}{\partial x_{\mu}} \,+\,
\frac{q}{2}\,\sigma_{\mu\nu}F^{\mu\nu}\! f_{\frac{1}{2}} -
\hspace{0.03cm}
m\hspace{0.02cm}\gamma_{\mu}\gamma_{5}\,\frac{\partial f_{0}}{\partial x_{\mu}} =
i\hspace{0.02cm}m \bigl(\hspace{0.03cm}\pi_{\mu}\gamma^{\mu}\gamma_{5} +\, m\hspace{0.02cm}\gamma_{5}\bigr)f_{1}.
\label{eq:9s}
\end{equation}
The differential equation for the even spinor $f_{0}$ is analyzed similar to Eq.\,(\ref{eq:9i}) by the replacement
\begin{equation}
f_{0} = \sqrt{\eta}\hspace{0.03cm}\varphi_{0},
\quad
\eta\equiv\frac{2\,}{m^2} \bigl[\hspace{0.03cm}\bar{f}_{0}(\gamma_{\mu}\pi^{\mu} - m)f_{0}\bigr].
\label{eq:9d}
\end{equation}
For the odd spinor $f_{\frac{1}{2}}$ we define an analogous replacement by introducing a new odd spinor $\theta_{\frac{1}{2}}$ by the rule
\begin{equation}
f_{\frac{1}{2}} \!= \sqrt{\eta}\,\theta_{\frac{1}{2}},
\label{eq:9f}
\end{equation}
with the {\it same} scalar function $\eta$ as it was defined in (\ref{eq:9d}). Taking into account the continuity equation in the form (\ref{eq:9o}) and replacement (\ref{eq:9f}), we obtain instead of (\ref{eq:9s})
\begin{equation}
\frac{1}{i}\,\pi_{\mu}\,\frac{\partial\hspace{0.02cm} \theta_{\frac{1}{2}}}{\partial x_{\mu}} \,+\,
\frac{q}{4}\,\sigma_{\mu\nu}F^{\mu\nu}\theta_{\frac{1}{2}} -
\hspace{0.02cm}m\hspace{0.02cm}\gamma_{\mu}\gamma_{5}\,\frac{1}{2\sqrt{\eta}}
\frac{\partial \sqrt{\eta}\hspace{0.02cm}\varphi_{0}}{\partial x_{\mu}} =
\frac{1}{2}\,i\hspace{0.02cm}m\bigl(\pi_{\mu}\gamma^{\mu}\gamma_{5} +\, m\hspace{0.02cm}\gamma_{5}\bigr)\varphi_{1},
\label{eq:9g}
\end{equation}
where on the right-hand side we have also set $f_{1}\equiv\sqrt{\eta}\hspace{0.03cm}\varphi_{1}$. The equation obtained can be related to the equation of motion of a spin in external field in the form (\ref{eq:1q}), but instead of the even spinor $\psi(\tau)$, here we have the odd one $\theta_{\frac{1}{2}}(\tau)$. The latter can be identified with the auxiliary Grassmann spinor $\theta(\tau)$ we have used throughout this work. Next, the spinor $\varphi_{1}(\tau)$ on the right-hand side of (\ref{eq:9g}) is even and it can be related to our commuting spinor $\psi(\tau)$ by simple setting
\[
\varphi_{1} \equiv \psi.
\]
\indent
The expression in parentheses on the right-hand side of (\ref{eq:9g}) should be considered as
Grassmann-odd by virtue of the property of being odd of the original operator expression which correlates with it (see the text after formula (\ref{eq:9p})). The Grassmann-odd parity of this expression can be made explicitly if we reintroduce the dimensionless Grassmann scalar $\hbar^{1/2}\chi$ as a multiplier. Taking into account all the above-mentioned and the relation $\dot{x}_{\mu} = \pi_{\mu}/m$, we obtain the final equation for the odd spinor $\theta_{\alpha}$:
\begin{equation}
\frac{1}{i}\,\frac{d\hspace{0.02cm}\theta}{d\hspace{0.01cm}\tau} \,+\,
\frac{q}{4\hspace{0.02cm}m}\,\sigma_{\mu\nu}F^{\mu\nu}\theta\, +\, \dotsb\, =
\hbar^{1/2}\biggl(\frac{i\hspace{0.02cm}m}{2}\,\chi\hspace{0.02cm}\dot{x}^{\mu}
\bigl(\gamma_{\mu}\gamma_{5}\psi\bigr) +\, \frac{i\hspace{0.02cm}m}{2}\,\chi\bigl(\gamma_{5}\hspace{0.02cm}\psi\bigr)\!\biggr).
\label{eq:9h}
\end{equation}
Here, the dots denote the contribution of the last term on the left-hand side of Eq.\,(\ref{eq:9g}). Its physical meaning is not clear. The terms on the right-hand side of (\ref{eq:9h}) can be obtained by varying with respect to $\bar{\theta}$ from the terms which must be added to the Lagrangian (\ref{eq:1t}):
\begin{equation}
L =\,\ldots\,+\,\hbar^{1/2\!}\hspace{0.01cm}\left\{\left(\frac{i\hspace{0.02cm}m}{2}\,
\chi\hspace{0.03cm}\dot{x}^{\mu}\bigl(\bar{\theta}\gamma_{\mu}\gamma_{5}\psi\bigr) +\, \frac{i\hspace{0.02cm}m}{2}\,\chi\bigl(\bar{\theta}\gamma_{5}\psi\bigr)\!\right)
+ (\mbox{conj.\,part})\right\}.
\label{eq:9j}
\end{equation}
Finally, under the mapping of the Lagrangian (\ref{eq:1t}) into (\ref{ap:A1}) the expressions in braces should be identified with the Grassmann pseudovector $\xi_{\mu}$ and pseudoscalar $\xi_5$
by the rule
\[
\begin{split}
%(\bar{\theta}\theta)\hspace{0.03cm}
\xi_{\mu} &\,\sim\, \hbar^{1/2}\bigl(\bar{\theta}\gamma_{\mu}\gamma_{5}\psi\bigr) \,+\, (\mbox{conj.\,part}),\\
%
%(\bar{\theta}\theta)\hspace{0.03cm}
\xi_{\hspace{0.02cm}5} &\,\sim\, \hbar^{1/2}\bigl(\bar{\theta}\gamma_{5}\psi\bigr) \,+\, (\mbox{conj.\,part}),
\end{split}
\]
and, thereby, we can obtain the missing terms (\ref{eq:9q}) in our map (for the proper time gauge $e = 1/m$).
Although we have obtained here the equation of motion for the odd spinor $\theta_{\alpha}$, Eq.\,(\ref{eq:9h}), a similar equation can be obtained for the even spinor $\psi_{\alpha}$ by changing Grassmann parity of the spinors $(f_{0}\!,\hspace{0.02cm} f_{1})$ and $(f_{\frac{1}{2}}\!,\hspace{0.02cm}f_{\frac{3}{2}})$ in the decomposition (\ref{eq:9a}) to the opposite one.

%%%%%%%%%%%%%%%%%%% Conclusion %%%%%%%%%%%%%%%%%%%%%%%

\section{Discussion}
\setcounter{equation}{0}
\label{section_12}

In the preceding section it was shown that to construct the map into a complete Lagrangian (\ref{ap:A1}) possessing $n = 1$ local proper-time supersymmetry, the initial Lagrangian (\ref{eq:1t}) must also possess some supersymmetry (fermion symmetry). To accomplish these ends, we must add the terms of the form (\ref{eq:9j}) to (\ref{eq:1t})  containing the auxiliary anticommuting spinor $\theta_{\alpha}$. Moreover, the obtained equation (\ref{eq:9h}) for the odd spinor serves as a hint that the spinor should be considered as an {\it independent} dynamical variable subject to own dynamical equation. And finally, this odd spinor $\theta_{\alpha}$ should be related to its superpartner -- the even spinor $\psi_{\alpha}$, and thus we have to consider a single {\it superspinor}: $\Theta_{\alpha} = \theta_{\alpha} + \eta\hspace{0.04cm}\psi_{\alpha}$,
as was done in the paper \cite{sorokin_1989}. Here, $\eta$ is a real odd scalar. The next step forward in this direction is to use at the outset all considered variables $(\psi_{\alpha}, \theta_{\alpha}, \xi_{\mu})$ for a description of the spin degrees of freedom, assuming them to be equivalent. This approach is known in literature as the construction of Lagrangians with {\it doubly supersymmetry}, i.e. possessing both the local (world-time) and the global (space-time) SUSY \cite{kowalski-glikman_1988, kowalski_1988}.\\
\indent
However, it is worth noting that the Lagrangians suggested in \cite{sorokin_1989, kowalski_1988} containing the superspinor $\Theta_{\alpha}$ as a variable, can hardly be considered as the desired extension of the Lagrangian (\ref{eq:1t}). The commuting spinor $\psi_{\alpha}$ in these models always plays a role of only auxiliary nondynamical quantity that is unacceptable for us. This circumstance was already mentioned in Introduction.\\
\indent
The Lagrangians suggested in the papers by Berkovits \cite{berkovits_2002, berkovits_2001} are closely related to the required generalization of the Lagrangian (\ref{eq:1t}). Berkovits' Lagrangians have been formulated within the framework of the so-called {\it pure spinor formalism} \cite{berkovits_2000} in which kappa (Siegel) symmetry is replaced by a BRST-like invariance, and describe the ten-dimensional superparticle coupling to a super-Maxwell or super-Yang-Mills background fields. One of the key point in this approach is introducing into consideration the commuting pure spinor ghost variable $\psi_{\alpha}$ (in our present notation) satisfying $\psi\hspace{0.02cm}\gamma^m \psi = 0$ for $m=1$ to $10$ along with the dynamical
anticommuting spinor $\theta_{\alpha}$. Thus, for example, the BRST invariant action with the first order Lagrangian for the $N = 1, \, D = 10$ superparticle in the super-Maxwell background is
\begin{equation}
S_{pure} = S_0 + S_{int},
\label{eq:12q}
\end{equation}
where
\[
S_0 = \!\int\!\!d\tau \hspace{0.02cm}\bigl(P_{m}\hspace{0.02cm}\dot{x}^{m} - \frac{1}{2}\hspace{0.035cm}P_{m}P^{m} + p_{\alpha}\hspace{0.02cm}\dot{\theta}^{\alpha} + \varphi_{\alpha}\hspace{0.01cm}\dot{\psi}^{\alpha}\bigr),
\]
\[
S_{int} =  q\!\int\!d\tau \hspace{0.02cm}\bigl[\hspace{0.02cm}\dot{\theta}^{\alpha} A_{\alpha}(x,\theta) + \Pi^{m\!}A_{m}(x,\theta) - i\hspace{0.02cm}d_{\alpha}W^{\alpha}(x,\theta) - i\hspace{0.02cm}(\varphi\hspace{0.02cm}\gamma^{mn}\psi) F_{mn}(x,\theta)\bigr].
\]
Here,
\[
\Pi^{m} \equiv \dot{x}^{m} + \frac{i}{2}\, (\theta\gamma^m\dot{\theta}), \quad
d_{\alpha} \equiv p_{\alpha} - \frac{i}{2}\,P_{m}(\gamma^m\theta)_{\alpha},
\]
$P_m$ is the canonical momentum for $x^m$, the commuting spinor $\varphi_{\alpha}$ is the canonical  momentum for $\psi_{\alpha}$ and the anticommuting spinor $p_{\alpha}$ is the canonical one for $\theta_{\alpha}$;
%$A_{\alpha}(x, \theta)$ and
%$A_m (x, \theta)$ are the spinor and vector super-Maxwell gauge superfields;
$F_{mn}(x, \theta)$ and $W_{\alpha}(x, \theta)$ are the super-Maxwell superfield strengths (see \cite{berkovits_2002} for the definition of the other notations). The equations of motion for the spinors $\psi_{\alpha}$ and $\theta_{\alpha}$ have the form
\begin{align}
i\hspace{0.05cm}\frac{d\hspace{0.015cm}\psi_{\alpha}}{d\hspace{0.02cm}\tau} &= -q\hspace{0.03cm}(\gamma^{mn}\psi)_{\alpha} \hspace{0.02cm}F_{mn}(x,\theta),&\label{eq:12w}\\
i\hspace{0.05cm}\frac{d\hspace{0.015cm}\theta_{\alpha}}{d\hspace{0.02cm}\tau} &= -q\hspace{0.025cm}W_{\alpha}(x,\theta).&\label{eq:12e}
\end{align}
The leading terms in the $\theta$-expansion of the superfields  $F_{mn}(x, \theta)$ and $W_{\alpha}(x,\theta)$ are
\begin{equation}
\begin{split}
F_{mn}(x, \theta) &= F_{mn}(x) + \bigl[\bigl(\theta\gamma_{n}\partial_{m}\chi(x)\bigr) -
\bigl(\theta\gamma_{m}\partial_{n}\chi(x)\bigr)\bigr] + \,\ldots\,,\\
W_{\alpha}(x,\theta) &= \chi_{\alpha}(x) - \frac{1}{4}\,(\gamma^{mn}\theta)_{\alpha}\hspace{0.02cm} F_{mn}(x) +\, \ldots\, .
\label{eq:12r}
\end{split}
\end{equation}
The lowest component of the superfield  $F_{mn}(x, \theta)$ is the vector field strength $F_{mn}(x)$ and the lowest component of $W_{\alpha}(x,\theta)$ is the spinor background field $\chi_{\alpha}(x)$. We see that equation (\ref{eq:1q}) is contained in (\ref{eq:12w}) as its integral part (in the action (\ref{eq:12q}) the parametrization gauge $e = - 1/2$ is chosen rather than the proper time one $e = 1/m$). Further, the second term in the expansion of $W_{\alpha}(x, \theta)$ by substitution into (\ref{eq:12e}), enables us to reproduce equation (\ref{eq:9h}). Notice that, although the action (\ref{eq:12q}) was suggested for the ten-dimensional superparticle, this approach can be also extended to the low-dimension pure-spinor superparticles \cite{berkovits_lect_2002, grassi_2005}, in particular, for $D = 4$ one. Thus the action (\ref{eq:12q}) is the best candidate for the desired extension of the action with Lagrangian (\ref{eq:1t}) (in the paper \cite{berkovits_2001} the action (\ref{eq:12q}) has been defined for the case of the interaction with a super-Yang-Mills background field, that correspondingly requires introducing the self-conjugate pair of the Grassmann color charges $(\theta^{\dagger i},\theta^i))$. However, here we are faced with different problem. In the action (\ref{eq:12q}) we have the interaction of a particle with the supersymmetric gauge field, whereas in (\ref{eq:1t}) the usual vector field is presented. The question arises whether it is possible to define analog of the action (\ref{eq:12q}) for nonsupersymmetric background field, for example, by simply setting  $\chi_{\alpha}(x) \equiv 0$ in (\ref{eq:12q}) and (\ref{eq:12r}).\\
\indent

\section*{\bf Acknowledgments}

This work was supported in part by the grants of the President of Russian Federation for the support of the leading scientific schools (NSh-3003.2014.2, NSh-5007.2014.9).

%%%%%%%%%%%%%%%%%%%%%%%%%%%%%%%%%%%%%%%%%%%%%%%%%

\begin{appendices}
\numberwithin{equation}{section}

\section{ Lagrangian of a spinning particle }
\label{appendix_A}

Here, for convenience of future references we write out a complete form of the Lagrangian for a spinning massive particle in external non-Abelian gauge field given in the paper \cite{balachandran_1977}. We also write out the local $n=1$ super-transformation under which this Lagrangian is invariant, the constrain equations and the equations of motion for dynamical variables.\\
\indent
The most general Lagrangian for a classical relativistic spin-$\frac{1}{2}$ particle moving in the background non-Abelian gauge field is (we put $c=1$ for the speed of light)
\begin{align}
L=L_{0}+L_{m}+L_{\theta},  &\label{ap:A1}
\end{align}
where
\begin{align}
&L_{0} = -\displaystyle\frac{1}{2\hspace{0.02cm}e}\,\dot{x}_{\mu\hspace{0.02cm}}\dot{x}^{\mu} - \displaystyle\frac{i}{2}\,\xi_{\mu\hspace{0.02cm}}\dot{\xi}^{\mu} + \displaystyle\frac{i}{2\hspace{0.02cm}e}\,\chi\hspace{0.02cm}\dot{x}_{\mu}\hspace{0.02cm}\xi^{\mu}, \label{ap:A2}\\
%\eqno{\rm (A.2)}
%
&L_{m} = -\displaystyle\frac{e}{2}\,m^2 + \displaystyle\frac{i}{2}\,\xi_{5\,}\dot{\xi_{5}} + \displaystyle\frac{i}{2}\,m\chi\hspace{0.02cm}\xi_{5}, \label{ap:A3}\\
&L_{\theta} = i\hbar\hspace{0.04cm}\theta^{\dagger{i\!}}D^{ij}\theta^{j}
+ \displaystyle\frac{i}{2}\,e\hspace{0.01cm}g\hspace{0.04cm}Q^{a\!}F^{a}_{\mu\nu}\hspace{0.035cm}
\xi^{\mu}\xi^{\nu}.  \label{ap:A4}
\end{align}
Here, $\xi_{\mu},\,\mu=0,1,2,3$, and $\xi_{5}$ are dynamical variables\footnote{\,Here, in contrast to \cite{balachandran_1977}, as the notation of spin variable we use the letter $\xi$ instead of generally accepted notation $\psi$, since the latter is used throughout the present work for the notation of bispinor $\psi_{\alpha}$.} describing the relativistic spin dynamics of the massive particle. These variables are elements of the Grassmann algebra \cite{berezin_1975}. The Lagrangian is invariant up to a total derivative under the following infinitesimal supersymmetry transformations
\begin{align}
&\delta{x}_{\mu}=i\hspace{0.02cm}\alpha\hspace{0.02cm}\xi_{\mu}, \notag\\
&\delta{\xi}_{\mu}=-\alpha\Bigl(\dot{x}_{\mu}-\frac{1}{2}\,i\hspace{0.01cm}\chi\hspace{0.02cm}
\xi_{\mu}\Bigr)\!\Big/e, \notag\\
&\delta{e}=-i\hspace{0.02cm}\alpha\chi,  \label{ap:A5}\\
&\delta{\chi}=2\hspace{0.02cm}\dot{\alpha}, \notag\\
&\delta{\xi}_{5}=m\hspace{0.02cm}\alpha, \notag\\
&\delta{\theta}^{i}=(g/\hbar)\hspace{0.02cm}\alpha\,\xi^{\mu\!}A_{\mu}^{a}(t^{a})^{ij}\theta^{j}, \notag
\end{align}
where $\alpha=\alpha(\tau)$ is an arbitrary Grassmann-valued function.\\
\indent
Varying the variables $e$, $\chi$ and $\xi_{5}$, we obtain the constraint equations
\begin{align}
&(\hspace{0.02cm}\dot{x}^2-i\chi\hspace{0.02cm}{\dot{x}}_{\mu\hspace{0.02cm}}\xi^{\mu})/e^{2}-m^{2\!} +\hspace{0.02cm} ig\hspace{0.025cm}Q^{a\!}F^{\hspace{0.005cm}a}_{\mu\nu}\hspace{0.035cm}\xi^{\mu}\xi^{\nu}=0, \notag\\
&\dot{x}_{\mu\hspace{0.02cm}}\xi^{\mu} + m\hspace{0.02cm}e\hspace{0.02cm}\xi_{5}=0, \label{ap:A6}\\
&2\hspace{0.03cm}\dot{\xi}_{5}-m\chi=0, \notag
\end{align}
which by the specific choice of the proper time gauge $e\!=\!1/m,\chi\!=\!0$ and $\xi_{5}\!=\!0$  are reduced to
\begin{align}
&m^2\dot{x}^{2}-m^2\!+ig\hspace{0.02cm}Q^{a\!}F^{a}_{\mu\nu}\hspace{0.035cm}
\xi^{\mu}\xi^{\nu}=0, \label{ap:A7}\\
&\dot{x}_{\mu\hspace{0.02cm}}\xi^{\mu}=0. \notag
\end{align}
Finally, variation over the remaining dynamical variables gives the equations of motion
\begin{align}
&\dot{\xi}_{\mu}-\frac{g}{m}\,Q^{a\!}F^{a}_{\mu\nu\,}\xi^{\nu}=0,  \label{ap:A8}\\
&\dot{\theta}^{i} + \frac{i\hspace{0.01cm}g}{\hbar}\hspace{0.02cm}\Bigl(A^{a}_{\mu\,}\dot{x}^{\mu}-\frac{i}{2m}\,
F^{a}_{\mu\nu}\,\xi^{\mu}\xi^{\nu}\Bigr)(t^{a})^{ij}\theta^j=0, \label{ap:A9}\\
&m\hspace{0.02cm}\ddot{x}_{\mu} - g\hspace{0.025cm}Q^{a\!}\Bigl(F^{a}_{\mu\nu\hspace{0.03cm}}\dot{x}^{\nu}-
\frac{i}{2\hspace{0.02cm}m}\,D^{ab}_{\mu}(x)F^{b}_{\nu\lambda}
\hspace{0.02cm}\xi^{\nu}\xi^{\lambda}\Bigr)=0.
\label{ap:A10}
\end{align}
Here, ${D}^{ab}_{\mu}(x)=\delta^{ab}\partial/\partial x^{\mu} + i\hspace{0.02cm}(g/\hbar)\hspace{0.01cm} A^{c}_{\mu}(x)(T^{c})^{ab}$  is the covariant derivative in the adjoint representation, where $(T^{c})^{ab}\equiv\! -i\hspace{0.01cm}f^{cab}$. In deriving (A.10) we have made use of the equation of motion for the commuting color charge $Q^a(\equiv\theta^{\dagger}t^a\theta)$
\begin{align}
\dot{Q}^{a} + \frac{i\hspace{0.01cm}g}{\hbar}\hspace{0.02cm}\Bigl(A^{b}_{\mu\hspace{0.02cm}}\dot{x}^{\mu}-
\frac{i}{2\hspace{0.02cm}m}\hspace{0.02cm}F^{b}_{\mu\nu}\,\xi^{\mu}\xi^{\nu}\Bigr)(T^b)^{ac}
Q^{c}=0.
&\label{ap:A11}
\end{align}
This equation follows from the equation of motion for the Grassmann color charge $\theta^i$, Eq.\,(\ref{ap:A9}).
\indent
The color current of the particle, which enters as the source into the equation of motion for the gauge field,
\[
D^{ab}_{\mu}(x)F^{b\hspace{0.025cm}\mu\nu}(x)=j^{a\hspace{0.03cm}\nu}(x),
\]
is
\begin{align}
j^{a\hspace{0.02cm}\mu}(x) = g\!\!\int\!\!{d}\tau\Bigl(Q^{a}\dot{x}^{\mu}
-i\hspace{0.03cm}\xi^{\mu}\xi^{\nu}\frac{1}{m}\hspace{0.03cm}{D}^{ab}_{\nu}(x)\hspace{0.03cm}
Q^b\Bigr)
\delta^{(4)}(x-x(\tau)).
&\label{ap:A12}
\end{align}
\indent
Besides the initial complete expression (\ref{ap:A1}) we need a somewhat reduced form of the Lagrangian. For this purpose we use the last constraint equation in (\ref{ap:A6}), which expresses the one-dimensional gravitino field $\chi$ in terms of the quantity $\xi_5$:
\begin{equation}
\chi = \frac{2}{m}\,\dot{\xi}_5.
\label{ap:A13}
\end{equation}
Substituting (\ref{ap:A13}) into (\ref{ap:A2}) and (\ref{ap:A3}) we obtain the required form of $L_0$ and $L_m$, respectively
\begin{align}
&L_{0} = -\displaystyle\frac{1}{2\hspace{0.02cm}e}\,\dot{x}_{\mu\hspace{0.02cm}}\dot{x}^{\mu} - \displaystyle\frac{i}{2}\,\xi_{\mu\hspace{0.02cm}}\dot{\xi}^{\mu} + \displaystyle\frac{i}{m\hspace{0.02cm}e}\,\dot{x}_{\mu}\hspace{0.02cm}\dot{\xi}_{5}\hspace{0.02cm}
\xi^{\mu}, \label{ap:A14}\\
%
%\equiv -\displaystyle\frac{1}{2\hspace{0.02cm}e}\,\bigl(\dot{x}_{\mu} -\frac{i}{m}\,\dot{\xi}_5\hspace{0.02cm}\xi_{\mu}\bigl)^2 \,- \, \displaystyle\frac{i}{2}\,\xi_{\mu\hspace{0.02cm}}\dot{\xi}^{\mu}, \label{ap:A2}\\
%
&L_{m} = -\displaystyle\frac{e}{2}\,m^2 - \displaystyle\frac{i}{2}\,\xi_{5\,}\dot{\xi_{5}}.
\label{ap:A15}
\end{align}
We note especially that after the elimination (\ref{ap:A13}), the kinetic term $\xi_{5\hspace{0.02cm}}\dot{\xi}_{5}$ in (\ref{ap:A3}) changes its sign to opposite one. The supersymmetry transformations (\ref{ap:A5}) take the form:
\begin{align}
&\delta{x}_{\mu}=i\hspace{0.02cm}\alpha\hspace{0.02cm}\xi_{\mu},  \label{ap:A16}\\
&\delta{\xi}_{\mu}=-\alpha\Bigl(\dot{x}_{\mu} -
\displaystyle\frac{i}{m}\,\dot{\xi}_{5}\hspace{0.02cm}\xi_{\mu}\Bigr)\!\Big/e, \label{ap:A17}\\
&\delta{e}=-\frac{2\hspace{0.015cm}i}{m}\,\hspace{0.02cm}\alpha\hspace{0.015cm}\dot{\xi}_5,  \label{ap:A18}\\
&\delta{\xi}_{5}=m\hspace{0.02cm}\alpha,  \label{ap:A19}\\
&\delta{\theta}^{i}=(g/\hbar)\hspace{0.02cm}\alpha\,\xi^{\mu\!}A_{\mu}^{a}(t^{a})^{ij}\theta^{j}. \label{ap:A20}
\end{align}

%%%%%%%%%%%%%%%%%%%%%%%%%%%%%%%%%%%%%%%%%%%%%%%% Appendix B %%%%%%%%%%%%%%%%%%%%%%%%%%%%%%%%%%%%%%%%%%%%%%%%

\section{\bf Spinor matrix algebra}
\label{appendix_B}
\numberwithin{equation}{section}

In this appendix we give some necessary formulae of the spinor matrix algebra, which are used in the text. The first basic formula is
$$
\gamma^{\mu}\gamma^{\nu} = {\rm I}\cdot g^{\mu\nu} + i\hspace{0.02cm}\sigma^{\mu\nu}, \quad \sigma^{\mu\nu}\equiv\frac{1}{2\hspace{0.02cm}i}\,
[\hspace{0.02cm}\gamma^{\mu},\gamma^{\nu}\hspace{0.02cm}],
$$
where ${\rm I}$ is the unity spinor matrix. We use the metric $g^{\mu\nu}={\rm diag}(1,-1,-1,-1)$. The useful identity is also
\begin{equation}
\sigma^{\mu\nu}\gamma_5 = \frac{1}{2\hspace{0.015cm}i}\,\epsilon^{\hspace{0.02cm}\mu\nu\lambda\sigma\!}
\sigma_{\lambda\sigma},
\label{ap:B1}
\end{equation}
where $\gamma_5\equiv i\hspace{0.015cm}\gamma^{0}\gamma^{1}\gamma^{2}\gamma^{3}$; $\epsilon^{\hspace{0.02cm}\mu\nu\lambda\sigma}$ is the totally antisymmetric tensor so that $\epsilon^{0123}= +1$.\\
\indent
The expansion of the product of the $\gamma$- and $\sigma$-matrices reads
\begin{equation}
\begin{split}
&\sigma^{\mu\nu}\gamma^{\lambda} = \frac{1}{i}\left(g^{\nu\lambda}\gamma^{\mu} - g^{\mu\lambda}\gamma^{\nu}\right) - \epsilon^{\hspace{0.02cm}\mu\nu\lambda\sigma}\gamma_{\sigma}\gamma_5,\\
&\gamma^{\lambda}\sigma^{\mu\nu} = \frac{1}{i}\left(g^{\mu\lambda}\gamma^{\nu} - g^{\nu\lambda}\gamma^{\mu}\right) - \epsilon^{\hspace{0.02cm}\mu\nu\lambda\sigma}\gamma_{\sigma}\gamma_5.
\end{split}
\label{ap:B2}
\end{equation}
The formula of the expansion for the product of two $\sigma$-matrices has the following form:
\begin{equation}
\sigma^{\mu\nu\!}\sigma^{\lambda\sigma} = {\rm I}\cdot(\hspace{0.03cm}g^{\mu\lambda}g^{\nu\sigma} - g^{\mu\sigma}g^{\nu\lambda}) +
\frac{1}{i}\left(\hspace{0.03cm}g^{\nu\lambda}\sigma^{\mu\sigma} - g^{\mu\lambda}\sigma^{\nu\sigma} - g^{\nu\sigma}\sigma^{\mu\lambda} + g^{\mu\sigma}\sigma^{\nu\lambda}\right)
+ \frac{1}{i}\,\epsilon^{\hspace{0.02cm}\mu\nu\lambda\sigma}\gamma_5.
\label{ap:B3}
\end{equation}
Finally, for the product of three $\sigma$-matrices we have
\begin{equation}
\sigma^{\rho\delta}\sigma^{\mu\nu\!}\sigma^{\lambda\sigma} =
\label{ap:B4}
\end{equation}
\[
=\frac{\,1}{\,i}\,\bigl\{
g^{\lambda\nu}\!\!\left(g^{\rho\mu}g^{\delta\sigma}\!-\!g^{\rho\sigma}g^{\delta\mu}\right)\!-\!
g^{\mu\lambda}\!\!\hspace{0.02cm}\left(g^{\rho\nu}g^{\delta\sigma}\!-\!g^{\rho\sigma}g^{\delta\nu}
\right)\!-\!
g^{\nu\sigma}\!\!\left(g^{\rho\mu}g^{\delta\lambda}\!-\!g^{\rho\lambda}g^{\delta\mu}\right)\!+\!
g^{\mu\sigma}\!\!\left(g^{\rho\nu}g^{\delta\lambda}\!-\!g^{\rho\lambda}g^{\delta\nu}\right)
\bigr\}\!\cdot\!{\rm I}
\]
\[
+ \left(g^{\lambda\rho}g^{\sigma\delta} - g^{\sigma\rho}g^{\lambda\delta}\right)\sigma^{\mu\nu} +  \left(g^{\mu\rho}g^{\nu\delta} - g^{\mu\delta}g^{\nu\rho}\right)\sigma^{\lambda\sigma} + \left(g^{\mu\lambda}g^{\nu\sigma} - g^{\mu\sigma}g^{\lambda\nu}\right)\sigma^{\rho\delta}
\]
\[
+ \left(g^{\lambda\delta}g^{\nu\rho} - g^{\lambda\rho}g^{\nu\delta}\right)\sigma^{\mu\sigma} +  \left(g^{\nu\delta}g^{\sigma\rho} - g^{\sigma\delta}g^{\nu\rho}\right)\sigma^{\mu\lambda} + \left(g^{\mu\rho}g^{\lambda\delta} - g^{\mu\delta}g^{\lambda\rho}\right)\sigma^{\sigma\nu}
\]
\[
+ \left(g^{\mu\rho}g^{\sigma\delta} - g^{\mu\delta}g^{\sigma\rho}\right)\sigma^{\nu\lambda} + \left(g^{\mu\lambda}g^{\nu\delta} - g^{\lambda\nu}g^{\mu\delta}\right)\sigma^{\rho\sigma}
+ \left(g^{\lambda\nu}g^{\rho\mu} - g^{\mu\lambda}g^{\rho\nu}\right)\sigma^{\delta\sigma}
\]
\[
+ \left(g^{\lambda\nu}g^{\delta\sigma} - g^{\nu\sigma}g^{\delta\lambda}\right)\sigma^{\rho\mu} + \left(g^{\nu\sigma}g^{\rho\lambda} - g^{\nu\lambda}g^{\rho\sigma}\right)\sigma^{\delta\mu}
+ \left(g^{\mu\sigma}g^{\rho\lambda} - g^{\mu\lambda}g^{\delta\sigma}\right)\sigma^{\rho\nu}
\]
\[
+ \left(g^{\mu\lambda}g^{\rho\sigma} - g^{\mu\sigma}g^{\rho\lambda}\right)\sigma^{\delta\nu} + \left(g^{\nu\sigma}g^{\mu\delta} - g^{\mu\sigma}g^{\nu\delta}\right)\sigma^{\rho\lambda}
+ \left(g^{\mu\sigma}g^{\rho\nu} - g^{\nu\sigma}g^{\rho\mu}\right)\sigma^{\delta\lambda}
\]
\[
- \,\bigl\{g^{\lambda\nu\!}\epsilon^{\hspace{0.02cm}\rho\delta\mu\sigma} - g^{\mu\lambda\!}\epsilon^{\hspace{0.02cm}\rho\delta\nu\sigma} - g^{\nu\sigma\!}\epsilon^{\hspace{0.02cm}\rho\delta\mu\lambda} + g^{\mu\sigma\!}\epsilon^{\hspace{0.02cm}\rho\delta\nu\lambda}
\bigr\}\gamma_5.
\hspace{4.4cm}
\]
It is worthy of special emphasis that in the expansion (\ref{ap:B4}) an explicit form of the last term is not uniquely defined by virtue of the fact that there are exist the identities relating the metric tensor $g^{\mu \nu}$ and the antisymmetric tensor $\epsilon^{\mu \nu \lambda \sigma}$:
\begin{equation}
\begin{split}
&g^{\lambda\nu\!}\epsilon^{\hspace{0.02cm}\rho\delta\mu\sigma} - g^{\mu\lambda\!}\epsilon^{\hspace{0.02cm}\rho\delta\nu\sigma} - g^{\nu\sigma\!}\epsilon^{\hspace{0.02cm}\rho\delta\mu\lambda} + g^{\mu\sigma\!}\epsilon^{\hspace{0.02cm}\rho\delta\nu\lambda}= \\
&g^{\rho\sigma\!}\epsilon^{\hspace{0.02cm}\mu\nu\lambda\delta} - g^{\lambda\rho\!}\epsilon^{\hspace{0.02cm}\mu\nu\sigma\delta} - g^{\sigma\delta\!}\epsilon^{\hspace{0.02cm}\mu\nu\lambda\rho} + g^{\lambda\delta\!}\epsilon^{\hspace{0.02cm}\mu\nu\sigma\rho}= \\
&g^{\mu\delta\!}\epsilon^{\hspace{0.02cm}\lambda\sigma\rho\nu} - g^{\rho\mu\!}\epsilon^{\hspace{0.02cm}\lambda\sigma\delta\nu} - g^{\delta\nu\!}\epsilon^{\hspace{0.02cm}\lambda\sigma\mu} + g^{\rho\nu\!}\epsilon^{\hspace{0.02cm}\lambda\sigma\rho\delta\mu}.
\end{split}
\label{ap:B5}
\end{equation}
These relations arise, for example, in calculating the following trace:
\[
{\rm Sp}(\sigma^{\mu\nu}\sigma^{\lambda\sigma}\sigma^{\rho\delta}\gamma_{5})
\]
by making use\footnote{\,Calculation of the trace for a product of six $\gamma$-matrices and $\gamma_5$, has been considered, for instance, in \cite{macfarlane_1966}. However, the ambiguity of representation for the trace has not been discussed there.} of the formula (\ref{ap:B1}). Another useful identity of such a kind \cite{shuryak_vainshtein_1982} is
\begin{equation}
\epsilon^{\hspace{0.02cm}\mu\beta\gamma\delta\!}g^{\alpha\nu} - \epsilon^{\hspace{0.02cm}\mu\gamma\delta\alpha\!}g^{\beta\nu} + \epsilon^{\hspace{0.02cm}\mu\delta\alpha\beta\!}g^{\gamma\nu} - \epsilon^{\hspace{0.02cm}\mu\alpha\beta\gamma\!}g^{\delta\nu} =
\epsilon^{\hspace{0.02cm}\alpha\beta\gamma\delta\!}g^{\mu\nu}.
\label{ap:B6}
\end{equation}

%%%%%%%%%%%%%%%%%%%%%%%%%%%%%%%%%%%%%%%%%%%%%%%%% Appendix C %%%%%%%%%%%%%%%%%%%%%%%%%%%%%%%%%%%%%%%%%%%%%%%%%

\section{\bf A complete list of the bilinear identities}
\label{appendix_C}
\numberwithin{equation}{section}

%\section*{\bf Appendix C}
%\setcounter{equation}{0}

Here, we give a complete list of all 15 sets of the bilinear relations. These relations are given in the same consequence as in the paper  \cite{kaempffer_1981}. Our list is introduced in such a form that it simultaneously covers both the commutative and non-commutative cases (here, $[\,, ]$ designates commutator, and $\{\,, \}$ is anticommutator). Such writing, in particular, enables us to define immediately the correct expressions for bilinear relations containing the derivatives of the currents $(S, V_{\mu},\! \,^{\ast}T_{\mu \nu}, A_{\mu}, P)$ (see Section \ref{section_3}). Note that some expressions, namely (\ref{ap:C3}), (\ref{ap:C7}), (\ref{ap:C8}), (\ref{ap:C10}) and (\ref{ap:C11}), can be introduced in somewhat different but equivalent forms. This fact is connected with ambiguity in calculating the trace
${\rm Sp}(\sigma^{\mu \nu} \sigma^{\lambda \sigma} \sigma^{\rho \delta} \gamma_5)$, as was mentioned in closing the previous appendix. The equivalence of the different representations can be directly proved with using the identities (\ref{ap:B5}) and (\ref{ap:B6}).
%
%%%%%%%%%%%%%%%%%%%%%%%%%%%%%%%%%%%%%%%%%%%%%%%%% (C.1)       SS %%%%%%%%%%%%%%%%%%%%%%%%%%%%%%%%%%%%%%%%%%%%%%%%%
\begin{flalign}
SS & = \frac{1}{4}\,SS - \frac{1}{4}\,PP - \frac{1}{4}\,V_{\mu}V^{\mu} - \frac{1}{4}\,A_{\mu}A^{\mu} + \frac{1}{8}\hspace{0.01cm}\,^{\ast}T_{\mu\nu}\!\,^{\ast}T^{\mu\nu}, &\label{ap:C1}
\end{flalign}
%
%%%%%%%%%%%%%%%%%%%%%%%%%%%%%%%%%%%%%%%%%%%%%%%%%
% (C.2)       SV^{\mu} %%%%%%%%%%%%%%%%%%%%%%%%%%%%%%%%%%%%%%%%%%%%%%%%%
%
\begin{flalign}
S\hspace{0.01cm}V^{\mu} &= \frac{1}{4}\,\{S,V^{\mu}\} + \frac{1}{4}\,[\hspace{0.03cm}P,A^{\mu}\hspace{0.02cm}] + \frac{1}{8}\,\epsilon^{\mu\nu\lambda\sigma}[\hspace{0.03cm}V_{\nu},^{\ast}\!T_{\lambda\sigma}] - \frac{1}{4}\,\{A_{\nu},^{\ast}\!T^{\mu\nu}\},&\label{ap:C2}
\end{flalign}
%
%%%%%%%%%%%%%%%%%%%%%%%%%%%%%%%%%%%%%%%%%%%%%%%%% (C.3)       S\,^{\ast}T^{\mu\nu} %%%%%%%%%%%%%%%%%%%%%%%%%%%%%%%%%%%%%%%%%%%%%%%%%
%
\begin{flalign}
S\hspace{0.01cm}\,^{\ast}T^{\mu\nu} = \,&\frac{1}{4}\,\{S,\!\,^{\ast}T^{\mu\nu}\} + \frac{1}{8}\,\epsilon^{\mu\nu\lambda\sigma}\{P,\!\,^{\ast}T_{\lambda\sigma}\} - \frac{1}{4}\,\{A^{\mu},V^{\nu}\} + \frac{1}{4}\,\{V^{\mu},A^{\nu}\} - \frac{1}{4}\,\epsilon^{\hspace{0.02cm}\mu\nu\lambda\sigma}V_{\lambda}V_{\sigma}  &\label{ap:C3}\\
-\,&\frac{1}{4}\;\epsilon^{\hspace{0.02cm}\mu\nu\lambda\sigma\!}A_{\lambda}A_{\sigma} - \frac{1}{4}\,\epsilon^{\mu\nu\lambda\sigma}\,^{\ast}T_{\lambda\rho}\!\,^{\ast}T^{\rho\;}_{\;\;\sigma}, &\notag
\end{flalign}
%
%%%%%%%%%%%%%%%%%%%%%%%%%%%%%%%%%%%%%%%%%%%%%%%%% (C.4)         SA^{\mu} %%%%%%%%%%%%%%%%%%%%%%%%%%%%%%%%%%%%%%%%%%%%%%%%%
%
\begin{flalign}
SA^{\mu} &= \frac{1}{4}\,\{S,A^{\mu}\} - \frac{1}{4}\,[\hspace{0.03cm}P,V^{\mu}\hspace{0.03cm}] + \frac{1}{8}\,\epsilon^{\hspace{0.02cm}\mu\nu\lambda\sigma}[\hspace{0.03cm}
A_{\nu},^{\ast}\!T_{\lambda\sigma}] + \frac{1}{4}\,\{V_{\nu},^{\ast}\!T^{\mu\nu}\},  &\label{ap:C4}
\end{flalign}
%
%%%%%%%%%%%%%%%%%%%%%%%%%%%%%%%%%%%%%%%%%%%%%%%%% (C.5)      SP %%%%%%%%%%%%%%%%%%%%%%%%%%%%%%%%%%%%%%%%%%%%%%%%%
%
\begin{flalign}
SP &= \frac{1}{4}\;\{S,P\} - \frac{1}{4}\;[\hspace{0.03cm}V_{\mu},A^{\mu}\hspace{0.02cm}] - \frac{1}{16}\,\epsilon^{\hspace{0.02cm}\mu\nu\lambda\sigma}\,^{\ast}T_{\mu\nu}
\!\,^{\ast}T_{\lambda\sigma}, &\label{ap:C5}
\end{flalign}
%
%%%%%%%%%%%%%%%%%%%%%%%%%%%%%%%%%%%%%%%%%%%%%%%%% (C.6)       V^{\mu}V^{\nu} %%%%%%%%%%%%%%%%%%%%%%%%%%%%%%%%%%%%%%%%%%%%%%%%%
%
\begin{flalign}
V^{\mu}V^{\nu} &= \frac{1}{4}\,\{V^{\mu},V^{\nu}\} -  \frac{1}{4}\,\{A^{\mu},A^{\nu}\} - \frac{1}{4}\,g^{\mu\nu}(SS + PP + V_{\lambda}V^{\lambda} - A_{\lambda}A^{\lambda} - \frac{1}{2}\hspace{0.01cm}\,^{\ast}T_{\lambda\sigma}\!\,^{\ast}T^{\lambda\sigma}) &\label{ap:C6}\\
&\,+ \frac{1}{8}\,\epsilon^{\hspace{0.02cm}\mu\nu\lambda\sigma}[\hspace{0.03cm}
S,\,^{\ast}T_{\lambda\sigma}] + \frac{1}{4}\,[\hspace{0.03cm}P,\,^{\ast}T^{\mu\nu}]
- \frac{1}{4}\,\epsilon^{\mu\nu\lambda\sigma}[\hspace{0.03cm}V_{\lambda},A_{\sigma}\hspace{0.02cm}] + \frac{1}{4}\hspace{0.01cm}\{\!\,^{\ast}T^{\mu\;}_{\;\;\lambda},\!\,^{\ast}T^{\lambda\nu}\}, \notag
\end{flalign}
%
%%%%%%%%%%%%%%%%%%%%%%%%%%%%%%%%%%%%%%%%%%%%%%%%% (C.7)       V^{\mu}\,^{\ast}T^{\nu\lambda} %%%%%%%%%%%%%%%%%%%%%%%%%%%%%%%%%%%%%%%%%%%%%%%%%
%
\begin{flalign}
V^{\mu}\hspace{0.04cm}^{\ast}T^{\nu\lambda} &= \!\hspace{0.015cm} \frac{1}{4}\,g^{\mu\lambda}\{S,A^{\nu}\} - \frac{1}{4}\,g^{\mu\nu}\{S,A^{\lambda}\} + \frac{1}{4}\,\epsilon^{\hspace{0.02cm}\mu\nu\lambda\sigma}[\hspace{0.03cm}S,V_{\sigma}
\hspace{0.03cm}]
&\label{ap:C7}\\
&+ \frac{1}{4}\,g^{\mu\lambda}\hspace{0.02cm}[\hspace{0.04cm}P\hspace{0.02cm},V^{\nu}
\hspace{0.03cm}] - \frac{1}{4}\,g^{\mu\nu}[\hspace{0.04cm}P\hspace{0.02cm},V^{\lambda}\hspace{0.03cm}] - \frac{1}{4}\,\epsilon^{\mu\nu\lambda\sigma}\{P,A_{\sigma}\} \notag \\
&+ \frac{1}{4}\,\{V^{\mu},^{\ast}\!T^{\nu\lambda}\} -  \frac{1}{4}\,\{V^{\lambda},^{\ast}\!T^{\mu\nu}\} + \frac{1}{4}\,\{V^{\nu},^{\ast}\!T^{\mu\lambda}\}
- \frac{1}{4}\,g^{\mu\nu}\{V_{\sigma},^{\ast}\!T^{\sigma\lambda}\} + \frac{1}{4}\,g^{\mu\lambda}\{V_{\sigma},^{\ast}\!T^{\sigma\nu}\} \notag \\
&- \frac{1}{8}\,\epsilon^{\hspace{0.02cm}\nu\lambda\sigma\rho}[\hspace{0.03cm}
A^{\mu},^{\ast}\!T_{\sigma\rho}] + \frac{1}{4}\,\epsilon^{\hspace{0.02cm}\mu\lambda\sigma\rho}\hspace{0.03cm}[A_{\sigma},^{\ast}
\!T_{\rho\;}^{\;\;\nu\hspace{0.01cm}}]
- \frac{1}{4}\,\epsilon^{\hspace{0.02cm}\mu\nu\sigma\rho}[\hspace{0.03cm}A_{\sigma},^{\ast}
\!T_{\rho\;}^{\;\;\lambda\hspace{0.01cm}}],\notag
\end{flalign}
%
%%%%%%%%%%%%%%%%%%%%%%%%%%%%%%%%%%%%%%%%%%%%%%%%% (C.8)        V^{\mu\!}A^{\nu} %%%%%%%%%%%%%%%%%%%%%%%%%%%%%%%%%%%%%%%%%%%%%%%%%
%
\begin{flalign}
V^{\mu\!}A^{\nu} =\, &\frac{1}{4}\,\{S,\,^{\ast}T^{\mu\nu}\} - \frac{1}{8}\,\epsilon^{\hspace{0.02cm}\mu\nu\lambda\sigma}\{P,\,^{\ast}T_{\lambda\sigma}\} - \frac{1}{4}\,g^{\mu\nu}[\hspace{0.04cm}S,P\hspace{0.04cm}]
+ \frac{1}{4}\,\epsilon^{\hspace{0.02cm}\mu\nu\lambda\sigma}V_{\lambda}V_{\sigma} - \frac{1}{4}\,\epsilon^{\hspace{0.02cm}\mu\nu\lambda\sigma\!}A_{\lambda}A_{\sigma} &\label{ap:C8}\\
+\, &\frac{1}{4}\,\Bigl(\{V^{\mu},A^{\nu}\} + \{V^{\nu},A^{\mu}\} - g^{\mu\nu}\{V_{\lambda},A^{\lambda}\}\Bigr) + \frac{1}{4}\,\Bigl(\epsilon^{\hspace{0.02cm}\nu\lambda\sigma\rho}
\,^{\ast}T^{\mu\;}_{\;\;\;\lambda}\,^{\ast}T_{\sigma\rho} +
\epsilon^{\hspace{0.02cm}\mu\lambda\sigma\rho}\,^{\ast}T_{\sigma\rho}
\,^{\ast}T^{\;\;\nu}_{\lambda\;}\Bigr),  \notag
\end{flalign}
%
%%%%%%%%%%%%%%%%%%%%%%%%%%%%%%%%%%%%%%%%%%%%%%%%% (C.9)           V^{\mu}P %%%%%%%%%%%%%%%%%%%%%%%%%%%%%%%%%%%%%%%%%%%%%%%%%
%
\begin{flalign}
V^{\mu\!}P &= \frac{1}{4}\,[\hspace{0.04cm}S,A^{\mu}\hspace{0.02cm}] + \frac{1}{4}\,\{P,V^{\mu}\} + \frac{1}{8}\,\epsilon^{\hspace{0.02cm}\mu\nu\lambda\sigma}\{A_{\nu},^{\ast}\!T_{\lambda\sigma}\} - \frac{1}{4}\,[\,V_{\nu},^{\ast}\!T^{\mu\nu}], &\label{ap:C9}
\end{flalign}
%
%%%%%%%%%%%%%%%%%%%%%%%%%%%%%%%%%%%%%%%%%%%%%%%%% (C.10)            \,^{\ast}T^{\mu\nu}\,^{\ast}T^{\lambda\sigma} %%%%%%%%%%%%%%%%%%%%%%%%%%%%%%%%%%%%%%%%%%%%%%%%%
%
\begin{flalign}
\,^{\ast}T^{\mu\nu}\,^{\ast}T^{\lambda\sigma} =\, &\frac{1}{4}\,\bigl(g^{\mu\lambda}g^{\nu\sigma} - g^{\mu\sigma}g^{\nu\lambda}\bigr) \Bigl(SS - PP + V_{\rho}V^{\rho} + A_{\rho}A^{\rho} + \frac{1}{2}\hspace{0.01cm}\,^{\ast}T_{\rho\delta}\,^{\ast}T^{\rho\delta}\Bigr) &\label{ap:C10}\\
- \,&\frac{1}{4}\,\epsilon^{\hspace{0.02cm}\mu\nu\lambda\sigma}\{S,P\} + \frac{1}{8}\,\bigl(\hspace{0.02cm}g^{\nu\sigma}\epsilon^{\rho\delta\mu\lambda} - g^{\mu\sigma}\epsilon^{\rho\delta\nu\lambda} + g^{\nu\lambda}\epsilon^{\rho\delta\mu\sigma} + g^{\mu\lambda}\epsilon^{\rho\delta\nu\sigma}\bigr)[\hspace{0.03cm}S,\,^{\ast}T_{\rho\delta}
\hspace{0.01cm}] \notag\\
+\, &\frac{1}{4}\,\bigl[\hspace{0.03cm}P,\bigl(g^{\nu\sigma}\,^{\ast}T^{\mu\lambda} - g^{\mu\sigma}\,^{\ast}T^{\nu\lambda} + g^{\nu\lambda}\,^{\ast}T^{\mu\sigma} + g^{\mu\lambda}\,^{\ast}T^{\nu\sigma}\bigr)\bigr] \notag
\end{flalign}
\[
\begin{split}
-\,\frac{1}{4}\,&\Bigl(g^{\mu\lambda}\{V^{\sigma}\!,V^{\nu}\} - g^{\nu\lambda}\{V^{\sigma}\!,V^{\mu}\} - g^{\mu\sigma}\{V^{\lambda}\!,V^{\nu}\} + g^{\nu\sigma}\{V^{\lambda}\!,V^{\mu}\}\Bigr)\\
-\,\frac{1}{4}\,&\Bigl(g^{\mu\lambda}\{A^{\sigma}\!,A^{\nu}\} - g^{\nu\lambda}\{A^{\sigma}\!,A^{\mu}\} - g^{\mu\sigma}\{A^{\lambda}\!,A^{\nu}\} + g^{\nu\sigma}\{A^{\lambda}\!,A^{\mu}\}\Bigr)\\
&-\frac{1}{4}\,\Bigl(\epsilon^{\hspace{0.02cm}\mu\lambda\sigma\rho}V_{\rho\,}A^{\nu} - \epsilon^{\hspace{0.02cm}\nu\lambda\sigma\rho}V_{\rho\,}A^{\mu}\Bigr) +
\frac{1}{4}\Bigl(\epsilon^{\hspace{0.02cm}\mu\nu\lambda\rho}V^{\sigma\!}A_{\rho} - \epsilon^{\hspace{0.02cm}\mu\nu\sigma\rho}V^{\lambda\!}A_{\rho}\Bigr) \\
&+\frac{1}{4}\,\Bigl(\epsilon^{\hspace{0.02cm}\mu\lambda\sigma\rho\!}A_{\rho}V^{\nu} - \epsilon^{\hspace{0.02cm}\nu\lambda\sigma\rho\!}A_{\rho}V^{\mu}\Bigr) -
\frac{1}{4}\Bigl(\epsilon^{\hspace{0.02cm}\mu\nu\lambda\rho\!}A^{\sigma}V_{\rho} - \epsilon^{\hspace{0.02cm}\mu\nu\sigma\rho\!}A^{\lambda}V_{\rho}\Bigr)\\
+\,\frac{1}{4}\,&\Bigl(\{\!\,^{\ast}T^{\mu\nu},\!\,^{\ast}T^{\lambda\sigma}\} + \{\!\,^{\ast}T^{\mu\lambda},\!\,^{\ast}T^{\nu\sigma}\} - \{\!\,^{\ast}T^{\mu\sigma},\!\,^{\ast}T^{\nu\lambda}\}\Bigr)\\
+\,\frac{1}{4}\,&\Bigl(g^{\mu\lambda}\{\!\,^{\ast}T^{\nu\rho},\!\,^{\ast}T_{\rho\;}^{\;\;\sigma}\} - g^{\mu\sigma}\{\!\,^{\ast}T^{\nu\rho},\!\,^{\ast}T_{\rho\;}^{\;\;\lambda}\} - g^{\nu\lambda}\{\!\,^{\ast}T^{\mu\rho},\!\,^{\ast}T_{\rho\;}^{\;\;\sigma}\} + g^{\nu\sigma}\{\!\,^{\ast}T^{\mu\rho},\!\,^{\ast}T_{\rho\;}^{\;\;\lambda}\}\Bigr),
\end{split}
\]
%
%%%%%%%%%%%%%%%%%%%%%%%%%%%%%%%%%%%%%%%%%%%%%%%%% (C.11)           \,^{\ast}T^{\mu\nu}A^{\lambda} %%%%%%%%%%%%%%%%%%%%%%%%%%%%%%%%%%%%%%%%%%%%%%%%%
%
\begin{flalign}
\,^{\ast}T^{\mu\nu\!}A^{\lambda} =\, &\frac{1}{4}\,g^{\mu\lambda}\{S,V^{\nu\!\hspace{0.02cm}}\} - \frac{1}{4}\,g^{\nu\lambda}\{S,V^{\mu}\} + \frac{1}{4}\,\epsilon^{\hspace{0.02cm}\mu\nu\lambda\sigma}\{P,V_{\sigma}\}
&\label{ap:C11}\\
+\, &\frac{1}{4}\,g^{\mu\lambda}\hspace{0.02cm}[\hspace{0.04cm}P\hspace{0.02cm},A^{\nu}
\hspace{0.03cm}] \hspace{0.02cm} - \frac{1}{4}\,g^{\nu\lambda}[\hspace{0.04cm}P\hspace{0.02cm},A^{\mu}\hspace{0.03cm}] \hspace{0.02cm} - \frac{1}{4}\,\epsilon^{\hspace{0.02cm}\mu\nu\lambda\sigma}[\hspace{0.03cm}S,A_{\sigma}
\hspace{0.01cm}] \notag\\
-\, &\frac{1}{4}\,\{A^{\mu},^{\ast}\!T^{\nu\lambda}\} + \frac{1}{4}\,\{A^{\lambda},^{\ast}\!T^{\mu\nu}\} + \frac{1}{4}\,\{A^{\nu},^{\ast}\!T^{\mu\lambda}\}
+ \frac{1}{4}\,g^{\nu\lambda}\{A_{\sigma},^{\ast}\!T^{\sigma\mu}\} - \frac{1}{4}\,g^{\mu\lambda}\{A_{\sigma},^{\ast}\!T^{\sigma\nu}\} \notag\\
-\, &\frac{1}{8}\,\epsilon^{\hspace{0.02cm}\mu\nu\sigma\rho}[\hspace{0.03cm}
V^{\lambda},^{\ast}\!T_{\sigma\rho}] + \frac{1}{4}\,\epsilon^{\hspace{0.02cm}\mu\lambda\sigma\rho}[\hspace{0.03cm}V_{\sigma},^{\ast}
\!T_{\rho\;}^{\;\;\nu}\hspace{0.01cm}]
- \frac{1}{4}\,\epsilon^{\hspace{0.02cm}\nu\lambda\sigma\rho}[\hspace{0.03cm}V_{\sigma},^{\ast}
\!T_{\rho\;}^{\;\;\mu}\hspace{0.01cm}],\notag
\end{flalign}
%
%%%%%%%%%%%%%%%%%%%%%%%%%%%%%%%%%%%%%%%%%%%%%%%%% (C.12)          \,^{\ast}T^{\mu\nu}P %%%%%%%%%%%%%%%%%%%%%%%%%%%%%%%%%%%%%%%%%%%%%%%%%
%
\begin{flalign}
\,^{\ast}T^{\mu\nu\!}P =\, &\frac{1}{4}\,\{P,\!\,^{\ast}T^{\mu\nu}\} - \frac{1}{8}\,\epsilon^{\hspace{0.02cm}\mu\nu\lambda\sigma}\{S,\!\,^{\ast}T_{\lambda\sigma}\} + \frac{1}{4}\,\epsilon^{\hspace{0.02cm}\mu\nu\lambda\sigma}\{V_{\lambda},A_{\sigma}\} - \frac{1}{4}\,[\hspace{0.04cm}V^{\mu},V^{\nu}\hspace{0.01cm}] - \frac{1}{4}\,[\hspace{0.04cm}A^{\mu},A^{\nu}\hspace{0.01cm}] &\label{ap:C12}\\
+\, &\frac{1}{4}\,[\,^{\ast}T^{\mu\lambda},\!\,^{\ast}T_{\lambda\;}^{\;\;\nu}],\notag
\end{flalign}
%
%%%%%%%%%%%%%%%%%%%%%%%%%%%%%%%%%%%%%%%%%%%%%%%%% (C.13)         A^{\mu\!}A^{\nu} %%%%%%%%%%%%%%%%%%%%%%%%%%%%%%%%%%%%%%%%%%%%%%%%%
%
\begin{flalign}
A^{\mu\!}A^{\nu} =\, &\frac{1}{4}\,\{A^{\mu},A^{\nu}\} -  \frac{1}{4}\,\{V^{\mu},V^{\nu}\} - \frac{1}{4}\,g^{\mu\nu}(SS + PP - V_{\lambda}V^{\lambda} + A_{\lambda}A^{\lambda} - \frac{1}{2}\hspace{0.01cm}\,^{\ast}T_{\lambda\sigma}\!\,^{\ast}T^{\lambda\sigma}) &\label{ap:C13}\\
+\, &\frac{1}{8}\,\epsilon^{\hspace{0.02cm}\mu\nu\lambda\sigma}[\hspace{0.03cm}
S,\,^{\ast}T_{\lambda\sigma}] + \frac{1}{4}\,[\hspace{0.03cm}P,\,^{\ast}T^{\mu\nu}\hspace{0.01cm}]
+ \frac{1}{4}\,\epsilon^{\hspace{0.02cm}\mu\nu\lambda\sigma}[\hspace{0.03cm}V_{\lambda},A_{\sigma}] + \frac{1}{4}\hspace{0.01cm}\{\!\,^{\ast}T^{\mu\;}_{\;\;\;\lambda},\!\,^{\ast}T^{\lambda\nu}\}, \notag
\end{flalign}
%
%%%%%%%%%%%%%%%%%%%%%%%%%%%%%%%%%%%%%%%%%%%%%%%%% (C.14)         A^{\mu}P %%%%%%%%%%%%%%%%%%%%%%%%%%%%%%%%%%%%%%%%%%%%%%%%%
%
\begin{flalign}
A^{\mu\!}P &= -\frac{1}{4}\,[\hspace{0.03cm}S,V^{\mu}\hspace{0.02cm}] + \frac{1}{4}\,\{P,A^{\mu}\} - \frac{1}{8}\,\epsilon^{\hspace{0.02cm}\mu\nu\lambda\sigma}\{V_{\nu},^{\ast}\!T_{\lambda\sigma}\} - \frac{1}{4}\,[\hspace{0.03cm}A_{\nu},^{\ast}\!T^{\mu\nu}\hspace{0.01cm}],
&\label{ap:C14}
\end{flalign}
%
%%%%%%%%%%%%%%%%%%%%%%%%%%%%%%%%%%%%%%%%%%%%%%%%% (C.15)       PP %%%%%%%%%%%%%%%%%%%%%%%%%%%%%%%%%%%%%%%%%%%%%%%%%
%
\begin{flalign}
PP &= \frac{1}{4}\,PP - \frac{1}{4}\,SS - \frac{1}{4}\,V_{\mu}V^{\mu} - \frac{1}{4}\,A_{\mu}A^{\mu} - \frac{1}{8}\hspace{0.01cm}\,^{\ast}T_{\mu\nu}\!\,^{\ast}T^{\mu\nu}. &\label{ap:C15}
\end{flalign}

%%%%%%%%%%%%%%%%%%%%%%%%%%%%%%%%%%%%%%%%%%%%%%%%% Appendix D %%%%%%%%%%%%%%%%%%%%%%%%%%%%%%%%%%%%%%%%%%%%%%%%%
%\newpage

\section{Mapping the kinetic term}
\setcounter{equation}{0}

In Section \ref{section_3} we have produced a set of identities containing the tensor variables and their derivatives. We have shown by the example of the `scalar' equation (see Eqs.\,(\ref{eq:3t}), (\ref{eq:3y})) that on the left-hand side of these identities one cannot collect the required expressions with the derivative of tensor variables $(\dot{S}, \dot{V}_{\mu}, \ldots)$ when $\dot{\theta}_{\alpha} \neq 0$. For this reason in Section \ref{section_3} we have restricted our attention only to the case when the auxiliary spinor $\theta_{\alpha}$ is independent of $\tau$. In this appendix we would like to show that the condition $\theta_{\alpha} = {\rm const.}$ is not needed in deriving the mapping of the kinetic term (\ref{eq:3q}) in a class of Majorana spinors.\\
\indent
Let us introduce two types of expansions for the spinor structures containing the derivatives of spinors. The first of them is
\begin{equation}
\hbar^{1/2}\hspace{0.02cm}(\dot{\bar{\theta}}_{\beta}\psi_{\alpha}) = \frac{1}{4}\,
\Bigl\{-i\hspace{0.02cm}\dot{S}_1\hspace{0.01cm}\delta_{\alpha\beta} + \dot{V}^{\mu}_1(\gamma_{\mu})_{\alpha\beta} - \frac{i}{2}\,^{\ast}\dot{T}^{\mu\nu}_1(\sigma_{\mu\nu}\gamma_{5})_{\alpha\beta} +
i\hspace{0.01cm}\dot{A}^{\mu}_1(\gamma_{\mu}\gamma_{5})_{\alpha\beta} + \dot{P}_1(\gamma_{5})_{\alpha\beta}\Bigr\},
\label{ap:D1}
\end{equation}
where $\dot{S}_1 \equiv (\dot{\bar{\theta}}\psi), \, \dot{V}_1^{\mu} \equiv i\hspace{0.02cm}(\dot{\bar{\theta}} \gamma^{\mu}\psi)$ etc., and the second is
\begin{equation}
\hbar^{1/2}\hspace{0.02cm}(\bar{\theta}_{\beta}\dot{\psi}_{\alpha}) = \frac{1}{4}\,
\Bigl\{-i\hspace{0.02cm}\dot{S}_2\hspace{0.02cm}\delta_{\alpha\beta} + \dot{V}^{\mu}_2(\gamma_{\mu})_{\alpha\beta} - \frac{i}{2}\,^{\ast}\dot{T}^{\mu\nu}_2(\sigma_{\mu\nu}\gamma_{5})_{\alpha\beta} +
i\hspace{0.01cm}\dot{A}^{\mu}_2(\gamma_{\mu}\gamma_{5})_{\alpha\beta} + \dot{P}_2(\gamma_{5})_{\alpha\beta}\Bigr\},
\label{ap:D2}
\end{equation}
where in turn, $\dot{S}_2 \equiv (\bar{\theta} \dot{\psi}), \, \dot{V}_2^{\mu} \equiv i(\bar{\theta} \gamma^{\mu} \dot{\psi})$ etc.. We specially note that
$(\dot{S}_{1,\hspace{0.03cm}2}, \dot{V}_{1,\hspace{0.03cm}2}^{\mu}, \!\,^{\ast}\dot{T}^{\mu\nu}_{1,\hspace{0.03cm}2}\ldots)$ are merely symbols and the dot over the tensor variables is not the derivative over $\tau$. Only the sum of such two expressions
\[
\dot{S} = \dot{S}_1 + \dot{S}_2, \quad \dot{V}^{\mu} = \dot{V}_1^{\mu} + \dot{V}_2^{\mu},\,\dots\,.
\]
will be the ``actual'' derivative. In terms of these quantities the right-hand side of the expression for the kinetic term (\ref{eq:3e}) is divided into two pieces
\begin{equation}
\hbar\hspace{0.02cm}(\bar{\theta}_{\rm M}\theta_{\rm M})\!\!\left[\left(\!\frac{d\bar{\psi}_{\rm M}}{d\tau}\,\psi_{\rm M}\!\right)\! -
\!\left(\!\bar{\psi}_{\rm M}\,\frac{d\psi_{\rm M}}{d\tau}\!\right)\right]\! =
\label{ap:D3}
\end{equation}
\[
\begin{split}
=\, &\frac{1}{2}\, \bigl\{S\dot{S}_1 + V_{\mu}\dot{V}_1^{\mu} - \frac{1}{2}\,^{\ast}T_{\mu\nu}\,^{\ast}\dot{T}^{\mu\nu}_1 - A_{\mu}\dot{A}^{\mu}_1 - P\dot{P}_1\bigr\}\\
+\, &\frac{1}{2}\, \bigl\{S\dot{S}_2 + V_{\mu}\dot{V}_2^{\mu} - \frac{1}{2}\,^{\ast}T_{\mu\nu}\,^{\ast}\dot{T}^{\mu\nu}_2 - A_{\mu}\dot{A}^{\mu}_2 - P\dot{P}_2\bigr\}.
\end{split}
\]
The systems of algebraic equations containing the functions $\dot{S}_{1,\hspace{0.03cm}2}, \dot{V}_{1,\hspace{0.03cm}2},\,\ldots\,$ are obtained by the scheme described in Section \ref{section_3}. First, it is necessary to multiply the expression (\ref{eq:2q}) by the expressions (\ref{ap:D1}) and (\ref{ap:D2}), respectively (by analogy with Eq.\,(\ref{eq:3r})) and then to perform a crossed contraction with different spinor structures of the type
$\delta_{\beta \gamma} \delta_{\delta \alpha},\, \delta_{\beta \gamma} (\gamma_5)_{\delta \alpha},\,\ldots\,$. It is not difficult to see that a system of equations containing the functions $\dot{S}_2,\dot{V}_2^{\mu},\ldots$ will be completely similar to the system (\ref{eq:3u}), (\ref{eq:3i}) with appropriate replacements $\dot{S} \rightarrow \dot{S}_2, \, \dot{V}^{\mu} \rightarrow \dot{V}_2^{\mu}$ etc.\\
\indent
For a system of equations containing $\dot{S}_1,\dot{V}_1^{\mu},\ldots$ the situation is somewhat involved. Here, the right-hand side of equations will be completely similar to the right-hand side of corresponding equations for the functions $\dot{S}_2,\dot{V}_2^{\mu},\ldots$. On the left-hand side the functions with the `derivatives' and without derivatives should be rearranged among themselves with no change of a sign. By this means, instead of (\ref{eq:3u}) and (\ref{eq:3i}), we will now have a system of identities
\[
\begin{split}
4\hspace{0.02cm}S\dot{S}_1 &=  S\dot{S}_1 - P\dot{P}_1 - \bigl(V_{\mu}\dot{V}^{\mu}_1 + A_{\mu}\dot{A}^{\mu}_1\bigr) + \frac{1}{2}\hspace{0.01cm}\,^{\ast}T_{\mu\nu}\!\,^{\ast}\dot{T}^{\mu\nu}_1, \\
V_{\mu}\dot{V}^{\mu}_1 &= - \bigl(S\dot{S}_1 + P\dot{P}_1\bigr) - \frac{1}{2}\,\bigl(V_{\mu}\dot{V}^{\mu}_1 - A_{\mu}\dot{A}^{\mu}_1\bigr),\\
A_{\mu}\dot{A}^{\mu}_1 &= - \bigl(S\dot{S}_1 + P\dot{P}_1\bigr) + \frac{1}{2}\,\bigl(V_{\mu}\dot{V}^{\mu}_1 - A_{\mu}\dot{A}^{\mu}_1\bigr),\\
4\hspace{0.01cm}P\dot{P}_1 &= - \bigl(S\dot{S}_1 - P\dot{P}_1\bigr) - \bigl(V_{\mu}\dot{V}^{\mu}_1 + A_{\mu}\dot{A}^{\mu}_1\bigr) - \frac{1}{2}\hspace{0.01cm}\,^{\ast}T_{\mu\nu}\!\,^{\ast}\dot{T}^{\mu\nu}_1,\\
\!\,^{\ast}T_{\mu\nu}\!\,^{\ast}&\dot{T}^{\mu\nu}_1 = 3\hspace{0.015cm}(S\dot{S}_1 - P\dot{P}_1) - \frac{1}{2}\hspace{0.01cm}\,^{\ast}T_{\mu\nu}\!\,^{\ast}\dot{T}^{\mu\nu}_1.
\end{split}
\]
Inspection of these five equations has shown that only three of them are independent (in contrast to a similar system for $\dot{S}_2,\dot{V}_2^{\mu},\ldots$, where we had dealt with only two independent equations (\ref{eq:3o}) and (\ref{eq:3p})). It is convenient to represent these equations in the following form:
\begin{align}
V_{\mu}\dot{V}^{\mu}_1 &= A_{\mu}\dot{A}^{\mu}_1, \notag\\
S\dot{S}_1 - P\dot{P}_1 &= \frac{1}{2}\hspace{0.01cm}\,^{\ast}T_{\mu\nu}\!\,^{\ast}\dot{T}^{\mu\nu}_1,&\label{ap:D4}\\
-\hspace{0.015cm}2\bigl(S\dot{S}_1 + P\dot{P}_1\bigr) &=  V_{\mu}\dot{V}^{\mu}_1 + A_{\mu}\dot{A}^{\mu}_1.\notag
\end{align}
It is easy to see that by virtue of these equations all the contribution in (\ref{ap:D3}) containing the functions $\dot{S}_1,\dot{V}_1^{\mu},\ldots$ vanishes and thus all terms containing $d{\theta}_{\alpha}/d\tau$ are completely excluded from considera\-tion for the Majorana spinors as it occurs on the left-hand side of Eq.\,(\ref{eq:3e}). This circumstance can be considered as a good test for the correctness of the equations under examination and of the approach as a whole.

\end{appendices}

%\newpage

\end{document}